\documentclass[9pt, twocolumn, twoside]{pnas-new}
\DeclareUnicodeCharacter{00A0}{ }
\usepackage{graphicx}
\usepackage{parskip}
\usepackage{comment}
\usepackage{afterpage}
\usepackage{breqn}
\usepackage{amsmath}
\usepackage{nameref}
\usepackage{blindtext}

\usepackage{color}

\templatetype{pnasresearcharticle} % Choose template 
\title{Habit learning supported by efficiently controlled network dynamics in naive macaque monkeys}

\author[a]{Karol P. Szymula}
\affil[a]{Department of Bioengineering, School of Engineering \& Applied Science, University of Pennsylvania, Philadelphia, PA 19104 USA}
\author[b]{Fabio Pasqualetti}
\affil[b]{Department of Mechanical Engineering, University of California, Riverside, CA 92521 USA}
\author[c]{Ann M. Graybiel}
\affil[c]{Department of Brain and Cognitive Sciences, Massachusetts Institute of Technology, Cambridge, MA 02139 USA}
\author[d,*]{Theresa M. Desrochers}
\affil[d]{Department of Neuroscience, Department of Psychiatry and Human Behavior, Robert J. and Nancy D. Carney Institute for Brain Science, Brown University, Providence RI 02912 USA}
\author[a,e,f,g,h,i,*]{Danielle S. Bassett}
\affil[e]{Department of Electrical \& Systems Engineering, School of Engineering \& Applied Science, University of Pennsylvania, Philadelphia, PA 19104 USA}
\affil[f]{Department of Physics \& Astronomy, College of Arts \& Sciences, University of Pennsylvania, Philadelphia, PA 19104 USA}
\affil[g]{Department of Neurology, Perelman School of Medicine, University of Pennsylvania, Philadelphia, PA 19104 USA}
\affil[h]{Department of Psychiatry, Perelman School of Medicine, University of Pennsylvania, Philadelphia, PA 19104 USA}
\affil[i]{Santa Fe Institute, Santa Fe, NM 87501 USA}
\affil[*]{These two authors contributed equally.}

\leadauthor{Szymula}  


\begin{abstract}
Primates display a marked ability to learn habits in uncertain and dynamic environments. The associated perceptions and actions of such habits engage distributed neural circuits. Yet, precisely how such circuits support the computations necessary for habit learning remain far from understood. Here we construct a formal theory of network energetics to account for how changes in brain state produce changes in sequential behavior. We exercise the theory in the context of multi-unit recordings spanning the caudate nucleus, prefrontal cortex, and frontal eyefields of female macaque monkeys engaged in 60-180 sessions of a free scan task that induces motor habits. The theory relies on the determination of effective connectivity between recording channels, and on the stipulation that a brain state is taken to be the trial-specific firing rate across those channels. The theory then predicts how much energy will be required to transition from one state into another, given the constraint that activity can spread solely through effective connections. Consistent with the theory's predictions, we observed smaller energy requirements for transitions between more similar and more complex trial saccade patterns, and for sessions characterized by less entropic selection of saccade patterns. Using a virtual lesioning approach, we demonstrate the resilience of the observed relationships between minimum control energy and behavior to significant disruptions in the inferred effective connectivity. Our theoretically principled approach to the study of habit learning paves the way for future efforts examining how behavior arises from changing patterns of activity in distributed neural circuitry.
\end{abstract}

\dates{This manuscript was compiled on \today}
\doi{\url{www.pnas.org/cgi/doi/10.1073/pnas.XXXXXXXXXX}}

\begin{document}

\maketitle
\thispagestyle{firststyle}
\ifthenelse{\boolean{shortarticle}}{\ifthenelse{\boolean{singlecolumn}}{\abscontentformatted}{\abscontent}}{}

\section*{Introduction}

In a complex ever-changing environment, both human and non-human primates survive by learning to balance the need to gather new knowledge with the utilization of existing knowledge \cite{costa2019subcortical,ebitz2018exploration}. The formation of habits can be viewed as a natural consequence of locally optimal trade-offs between exploration and exploitation \cite{desrochers2010optimal}. The underlying cognitive processes may follow reinforcement learning algorithms \cite{kaelbling1996reinforcement}, in which the sampling of actions and the uncertainty of their outcomes inform decisions regarding exploration of new actions or exploitation of old ones \cite{gershman2018deconstrucing}. The brain mechanisms supporting such processes engage a distributed set of regions spanning the caudate nucleus associated with repetitive and stereotyped actions \cite{desrochers2015habit}, the ventral striatum and amygdala associated with reward and motivation \cite{costa2019subcortical}, and the prefrontal cortex associated with cognitive control \cite{ebitz2018exploration}.

Yet, precisely how this constellation of brain regions supports the computations necessary for habits to emerge remains far from understood. Recent efforts suggest that network approaches \cite{mitchell2011complexity} provide useful explanations for how cognitive processes arise from interacting brain regions \cite{rosenberg2020functional}. From intelligence to cognitive control, and from motivation to learning, disparate circuits are engaged that allow coordinated information processing and transmission \cite{barbey2018network,girn2019linking,bassett2017network}. The study of circuit engagement and function can be formalized in the language of network science \cite{bassett2018on}, and initial evidence suggests that individual differences in the pattern of inter-regional interactions track individual differences in exploratory behaviors and decision-making \cite{kao2019functional}, plasticity \cite{gallen2019brain}, reinforcement learning \cite{gerraty2018dynamic}, and skill learning \cite{bassett2015learning}. Although network approaches manifest striking face validity, the level of explanation has thus far been largely correlative \cite{bertolero2019on}. Continued progress will require a formal model positing and validating the network mechanisms of brain-behavior relations in habit formation.

Here we address this challenge by building upon and extending emerging work in the field of network control theory \cite{motter2015network,pasquletti2014controllability,liu2011controllability}. In the context of neural systems, the approach defines a state of the network to be the vector of regional (or cellular) activation. The theory then posits that the sequence of states is constrained by the energy required to transmute one state into another allowing activity to spread solely through known inter-regional links \cite{kim2019linear}. The theoretical background is particularly well-developed for linear systems \cite{kailath1980linear}, or linearizations of nonlinear systems \cite{kim2019linear,motter2015network}. In addition to predicting the effects of exogenous control signals such as electrical stimulation \cite{stiso2019white}, network control theory has also proven useful in accounting for the intrinsic capacity for cognitive control \cite{cornblath2019sex} and the contribution of single neurons to large-scale behaviors \cite{yan2017network}. We extend the approach in two ways. First, prior studies stipulated that activity could only flow along known structural links between regions; here, we instead allow inter-regional links to reflect effective connections \cite{friston2011functional}, which have recently been associated with short-term plasticity and learning \cite{dima2015neuroticism,buchel1999predictive}. Second, prior studies estimated the energy of brain state transitions independently from behavior; here, we instead explicitly posit (and validate) the notion that low energy state transitions characterize processes that are less cognitively demanding, as well as their associated behaviors \cite{braun2019brain}. 

We evaluate the theory in the context of multi-unit recordings spanning the caudate nucleus and prefrontal cortex of two macaque monkeys as they engage in 60-180 sessions of task performance inducing motor habits in the form of saccadic patterns \cite{desrochers2010optimal}. Acknowledging the pivotal nature of sequence-level strategies, we examine a task (\hyperref[fig1]{Figure 1}) in which the monkey must saccade among nine green dots (referred to as targets) on a rectangular grid in search for a baited target, which varied from trial to trial, was randomly selected using a pseudorandom schedule, and was visibly not distinguishable from the other targets during a task trial \cite{desrochers2015habit}; the ideal habit would be a sequence of saccades that spanned all nine target dots in a minimal time period. We define a brain state to be a vector of firing rates across recording channels. Further, we construct a neural network whose nodes are channels and whose edges are the strength of effective connectivity between channel firing rate time series; we represent the network as a weighted directed adjacency matrix. 

Our primary hypothesis is that pairwise differences in sequential behaviors during habit formation can be explained by the energy requirements of the accompanying neural state transitions. To interpret behavior, we represent saccade patterns as graphs that can be decomposed into 1D time series (\hyperref[fig1]{Figure 2A}), whose shape can be studied and whose complexity can be quantified. We observe that preferred saccade patterns change as a function of learning. Using network control theory, we compute the minimum control energy required to transition between chronologically ordered trial brain states and observe that energy decreases with learning. Finally, we show that the energy of state transitions predicts behavior in three distinct ways: (i) transitioning between more similar saccades patterns requires less energy, (ii) transitioning between more complex saccades patterns requires less energy, and (iii) a more organized, less-entropic selection of patterns during sessions requires less energy. This pattern of results is markedly consistent with the principles of maximum entropy, which have previously been shown to explain other features of neural and behavioral dynamics \cite{savin2017maximum,granot2013stimulus,meshulam2017collective,friston2006free,ortega2013thermodynamics}. Taken together, our work represents a theoretically principled study of habit learning that accurately predicts transitions in behavior from the energetics of transitions in neural states. 

\setlength{\parindent}{5ex}

\section*{Results}

\begin{figure}[h]
\centering
\includegraphics[width=\linewidth]{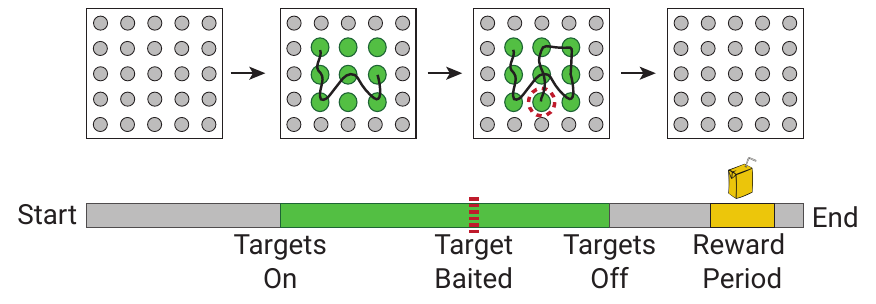}
\caption{\textbf{The Free Scan Task}. A visual depiction of a single trial from the free scan task. The trial begins (\emph{Start}) with a grid of small grey dots all equally sized and spaced. A 3$\times$3 grid of larger green targets replaces the central part of the gray grid (\emph{Targets On}). The monkey is allowed to freely scan the space of green targets. At a variable time unknown to the monkey, a target is baited (\emph{Target Baited}). For visualization purposes in the context of this exposition, the baited target is depicted surrounded by a red dashed circle; note that the outline is not present during the actual task. When the monkey’s gaze enters the baited target, the grid of green targets disappears and is replaced by all gray targets (\emph{Targets Off}). After a brief variable delay, a liquid reward is given to the monkey (\emph{Reward Period}) after which the trial is considered finished (\emph{End}).}
\label{fig1}
\end{figure}

The data analyzed in this work consist of behavioral measurements and neural recordings from two female macaque monkeys, Monkey G (MG) and Monkey Y (MY), while performing a free-view scanning task. All data were previously collected and reported in \cite{desrochers2015habit, desrochers2010optimal}. As depicted in \hyperref[fig1]{Figure 1}, during the task the monkey is shown a 3$\times$3 grid of green targets on a screen and allowed to visually navigate the grid-space freely. At a variable time, one of the targets is baited without the monkey's knowledge. When the monkey's gaze enters the baited target space, the green grid is replaced by smaller gray circles (marking the end of the task trial) and the monkey receives a reward after a short delay. 

\subsubsection*{Classification of Trial Representative Saccade Patterns}
\begin{figure*}[!t]
\includegraphics[width=\textwidth]{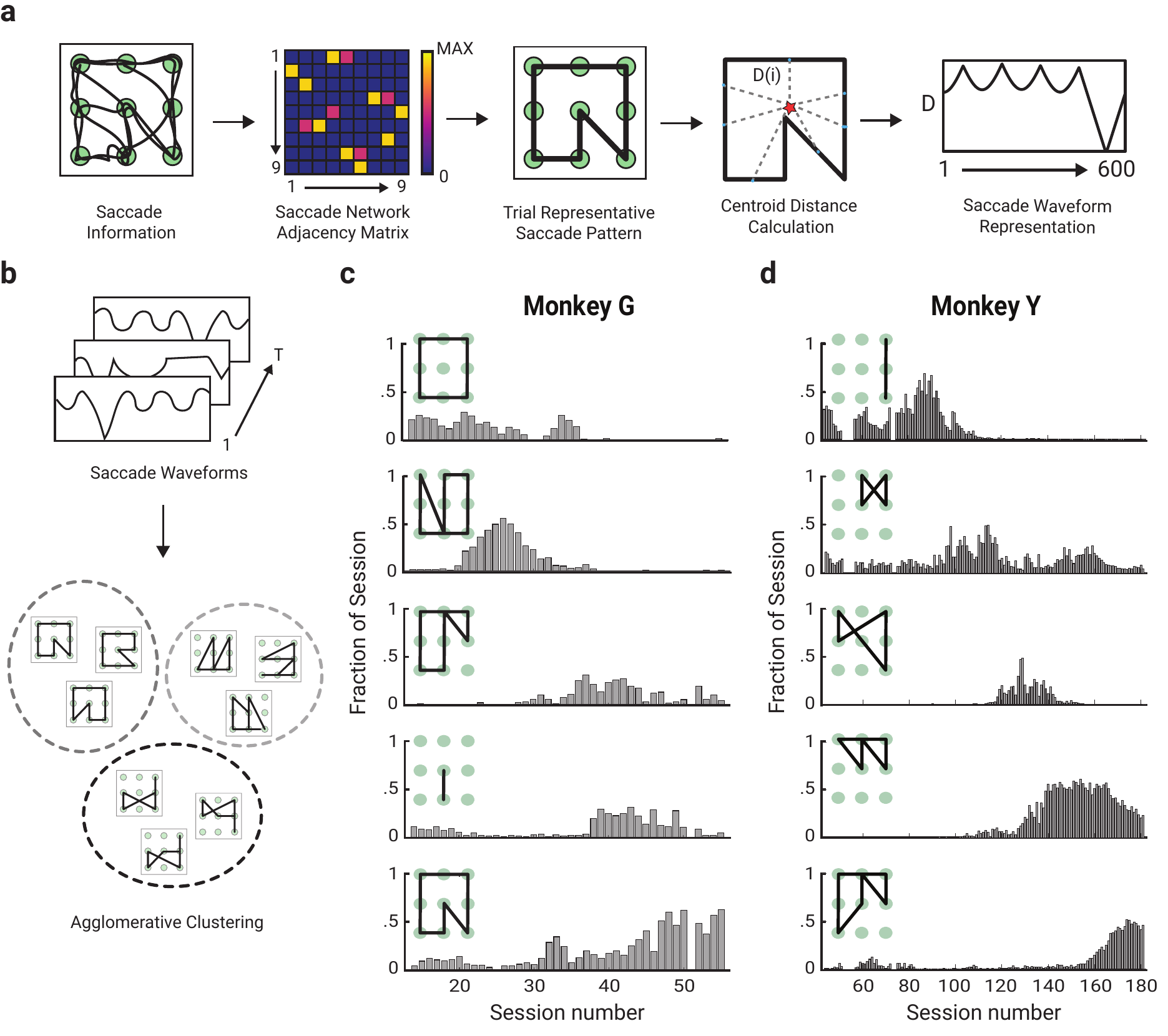}
\caption{\textbf{Classification of Trial Representative Saccade Patterns}. \textbf{(a)} Saccade information in the form of identified saccadic movements during a trial is collectively represented as an adjacency matrix, which in turn encodes a directed and weighted network. A total of nine nodes exist: one for every green target on the task grid. Edge weights are calculated as the number of times that an individual saccade is made from one node to another. The network is converted into a trial representative saccade pattern (TRSP) by identifying the network cycle with the greatest sum of edge weights along its path. Each TRSP is treated as a 2-D polygon in the task grid space consisting of a set of (x,y) points. The saccade waveform is taken to be the vector of Euclidean distances between the polygon centroid and all of its points. A 1-D interpolation is performed to reduce each saccade waveform to 600 values. \textbf{(b)} A dissimilarity matrix is constructed utilizing the saccade waveforms from all observed trials. Each element in the dissimilarity matrix is the Euclidean distance between two saccade waveforms. The dissimilarity matrix is of size T$\times$T where $T$ is the total number of trials for a given monkey. Agglomerative clustering with a threshold inconsistency coefficient of 0.95 was performed using the dissimilarity matrix to cluster all trial saccade waveforms. For Monkey G, we identified a total of 136 cluster whereas for Monkey Y, we identified a total of 346 clusters. \textbf{(c,d)} The five most prevalent cluster saccade patterns across all sessions are shown for each monkey. The saccade pattern shown is the one which is most similar to all other saccade patterns in the same cluster. }
\label{fig2}
\end{figure*}

Behavioral measurements were analyzed in the form of trial-specific chronological saccade sequences performed by a monkey during the task. We use the phrase \emph{individual saccade} to refer to a rapid eye movement from one target to another on the task grid. The phrase \emph{saccade sequence} then refers to a series of individual saccades that are performed one after another. Direct qualitative or quantitative comparison between the trial-specific saccade sequences is difficult due to variability in trial length and, as a result, the number of saccades per trial. Therefore, we first set out to arrange the list of individual saccades into a format that allows for interpretable comparison between trials. We began by converting each trial's saccade sequence into an adjacency matrix of a directed and weighted graph, $G\left(N,E\right)$, where $N$ is the number of nodes (one for each green grid target) and $E$ is the set of all edges that exist between nodes (\hyperref[fig2]{Figure 2A}). An edge between two nodes exists if a saccade was observed between the two specific grid targets. Furthermore, the weight of each edge is the total number of times the specific saccade is performed during the trial. We refer to this representation of the trial saccades as the \emph{saccade network}. 

Saccade patterns that create \emph{loops} (or sequences that start and end on the same target) are the most effective strategies since they allow for an efficient and organized approach to scanning the 3$\times$3 grid space \cite{desrochers2010optimal}. Accordingly, we identified the most observed cyclic saccade pattern in each trial by leveraging the concept of network paths (see \hyperref[subsubsec:methods:trsp]{Methods}). A series of edges that is traversed to move from one node in the network to another is called a path. In a saccade network, a path represents a set of observed saccades performed one after another. If the path's start node is the same as the path's end node then the path is cyclic and the represented saccade sequence is a loop. For each trial, we therefore defined the trial representative saccade pattern (TRSP) to be the cyclic path in the trial saccade network with the greatest sum of edge weights (\hyperref[fig2]{Figure 2A}). Intuitively, the TRSP is the cyclic sequence of saccades that was performed most frequently during a trial; see \textbf{Supplementary Figure 1} for a graphical depiction of TRSP identification.

Methods for computing the similarity between 1-D signals are numerous, easy to implement computationally, and provide simple intuitive understanding. Therefore, to make the comparison between trial representative saccade patterns as simple as possible we converted each pattern into a one-dimensional saccade waveform (\hyperref[fig2]{Figure 2A}). This dimensionality reduction step was made possible by representing the identified trial representative saccade pattern as a series of (x, y) points in the space of the task grid and calculating each points’ Euclidean distance from the centroid of all points. We take the similarity between any two trial saccade patterns to be a metric based on the Euclidean distance between the trial saccade waveforms (see \hyperref[subsubsec:methods:asf]{Methods}). We use this metric to group all the trials into clusters of saccade patterns based on their similarity to each other. Since the exact number of present clusters in the data was not known, the agglomerative clustering algorithm was used to group trial representative saccade patterns (\hyperref[fig2]{Figure 2B}). This algorithm starts by treating each object (i.e. saccade pattern) as a single cluster and uses an iterative process to merge pairs of objects into clusters until all objects are grouped into one large cluster. The output of the algorithm is a dendrogram (cluster tree) which depicts the order in which objects should be grouped during clustering. \par

In order to capture more natural divisions of our data during clustering, we used the inconsistency coefficient which is a useful metric in agglomerative clustering that compares the height of a link in a cluster tree to heights of all the other links underneath it in the tree. A small coefficient denotes little difference between the objects being grouped together, thereby suggesting that the clustering solution is a good fit to the data. Setting a threshold on the inconsistency coefficient during clustering enables the identified groupings to more closely represent the natural divisions found in the data. Furthermore, tuning the inconsistency coefficient threshold allows for optimization of the clusters without arbitrarily selecting a range of the maximum number of clusters to test. Using the elbow-method and the average within cluster sum-of-squares from a range of inconsistency coefficient thresholds, we selected an inconsistency coefficient of 0.95 to be the optimal threshold criterion for clustering. As a result, a total of 136 clusters were identified for Monkey G and 346 for Monkey Y; see \textbf{Supplementary Figures 2 and 3} for a full tabulation. In \hyperref[fig2]{Figure 2C}, the five most prominent trial representative saccade patterns for each monkey as well as their appearance frequency distribution across all sessions are shown. These patterns and their dynamics closely resemble those previously reported in Ref. \cite{desrochers2010optimal}. Both monkeys demonstrate non-uniform distributions of cluster appearance frequencies across sessions. Each of the main cluster types is acquired, preferentially performed, and dropped at varying time windows throughout the task. \par

\subsubsection*{Inferring Effective Connectivity}

\begin{figure}[!bp]
\includegraphics[width=\linewidth]{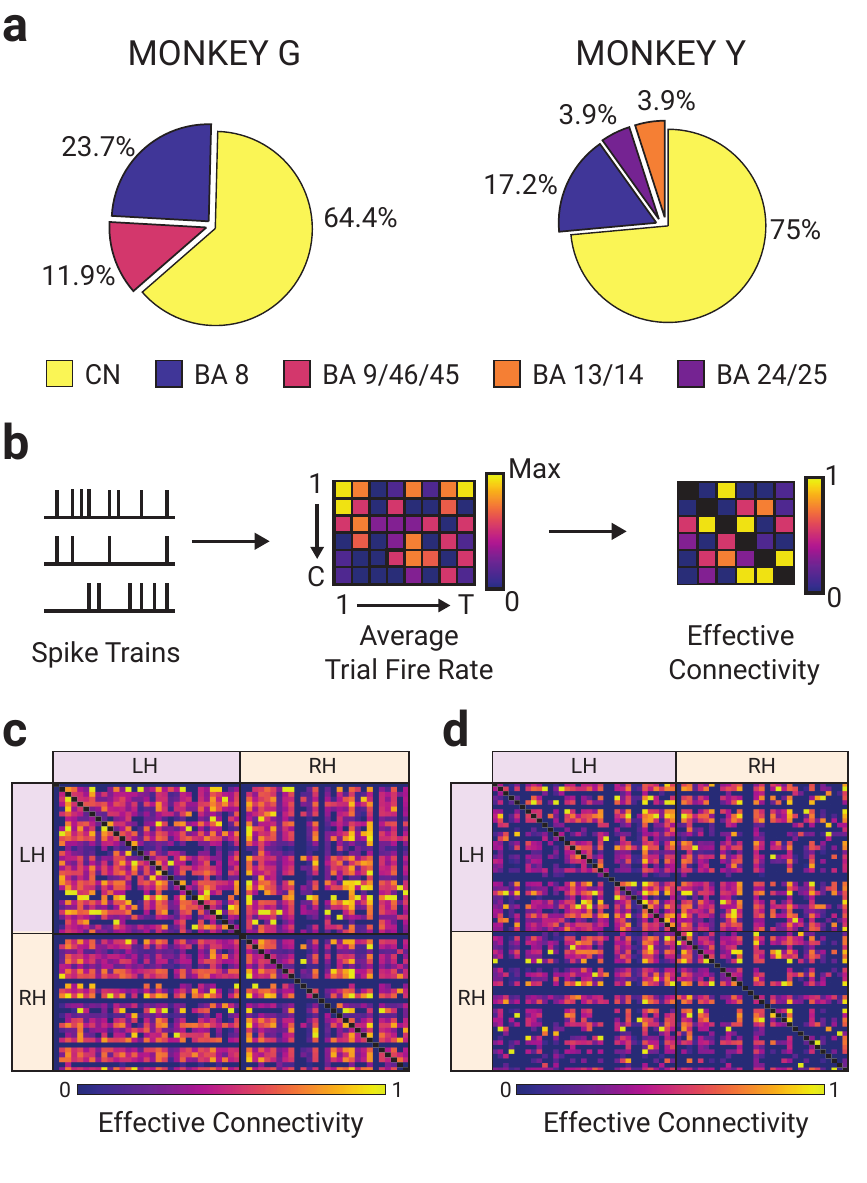}
\caption{\textbf{Inferring Effective Connectivity from Neural Activity}. \textbf{(a)} Summary of channel recording regions for both monkeys. Percentages denote the percent of all available channels which record from a given region across all trials. CN = caudate nucleus; BA 8, 9, 13-14, 24-25, 45-46 = Brodmann areas 8 (frontal eye fields), 9 (dorsolateral and medial prefrontal cortex), 13-14 (insula and ventromedial prefrontal cortex), 24-25 (anterior and subgenual cingulate), and 45-46 (\emph{pars triangularis} and middle frontal area). \textbf{(b)} Spike trains from all channels for a given session were converted into an average trial firing rate matrix. The matrix is of size $C\times T$, where $T$ is the number of trials for a session and $C$ is the number of available channels during the session. We used transfer entropy \cite{vicente2011transfer} to estimate the effective connectivity between session channels. \textbf{(c,d)} The overall combined effective connectivity matrices for both monkeys are shown where channels are organized according to their respective hemispheres (LH = Left; RH = Right). Both matrices are individually normalized by dividing all elements by the magnitude of the largest magnitude element.}
\label{fig3}
\end{figure}

After quantitatively characterizing behavior, our next aim was to demonstrate that pairwise differences in sequential saccade patterns during habit formation can be explained by the energy requirements of the accompanying neural state transitions. We approached the problem by using and extending recent advances in network control theory \cite{motter2015network,pasquletti2014controllability,liu2011controllability}. Fundamental to any control energy analysis is knowledge of the network structure and dynamics. Thus, as a first step we extract a network of interactions between the observed regions from available channels (\hyperref[fig3]{Figure 3A}). In both monkeys, more than half of the present channels were associated with the caudate nucleus (CN) and recordings from Brodmann area 8 (BA-8) were available in both. Although the anatomical location of each channel was known, no information regarding their anatomical connectivity was available and we therefore turned to alternative inference approaches \cite{friston2011functional}. 

Specifically, we inferred the effective connectivity of the regions using their neural activity (see \hyperref[subsec:methods:ec]{Methods}). For each session, the activity of the available channels was calculated as the average firing rate during each individual trial. This set of trial firing rates was used to calculate the transfer entropy \cite{vicente2011transfer} between all pairs of available channels, which provides basic structural information about the effective connectivity between them (\hyperref[fig3]{Figure 3B}). The effective connectivity matrices for Monkey G and Monkey Y are displayed in (\hyperref[fig3]{Figure 3C, 3D}).  

\subsubsection*{Assessing the Control Energy Required for Neural State Transitions}

\begin{figure*}[!t]
\includegraphics[width=\textwidth]{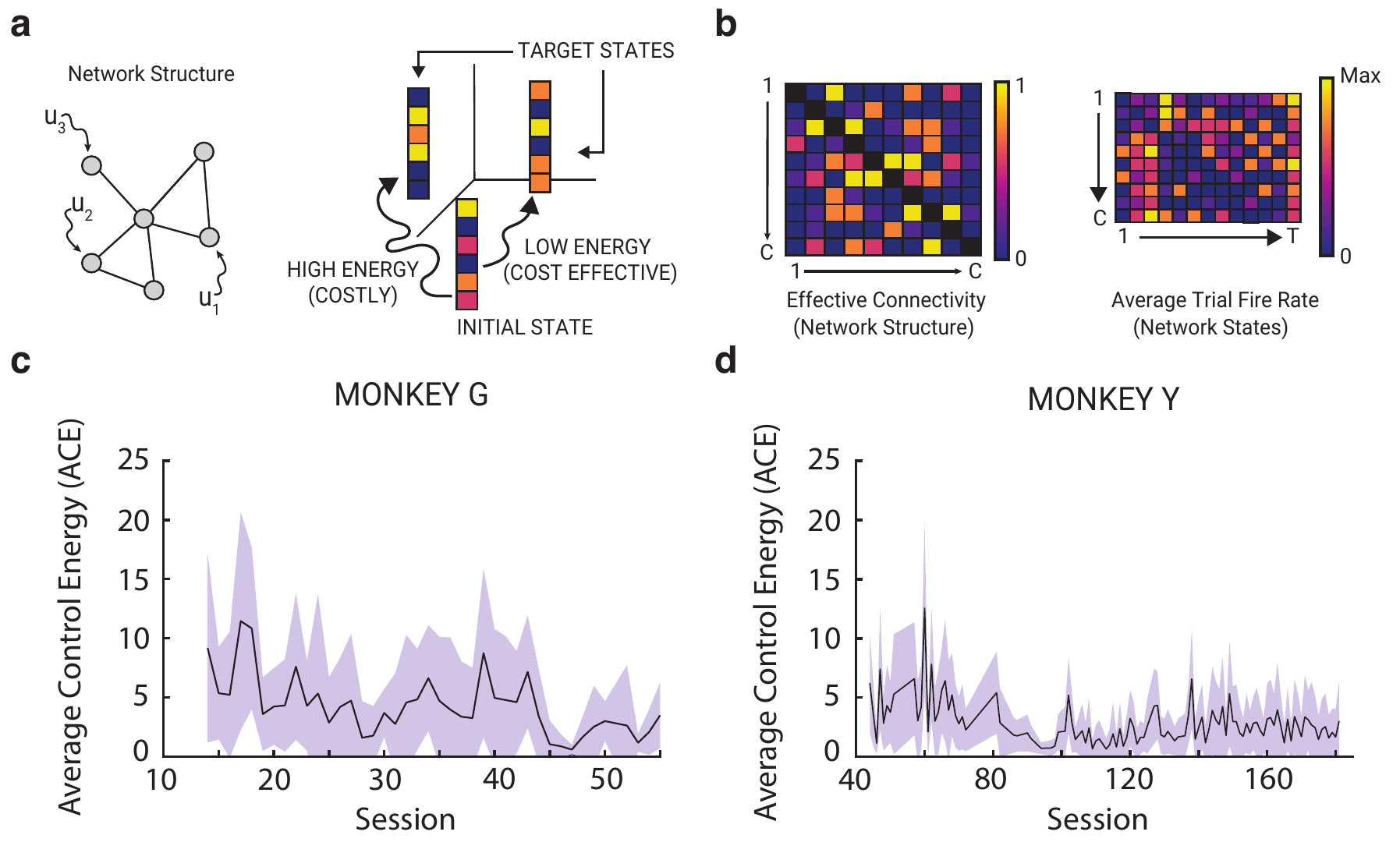}
\caption{\textbf{Estimating the Minimum Control Energy to Transition between Neural States}. \textbf{(a)} Visual depiction of control energy analysis. Given a network, the goal is to identify a set of time-dependent inputs (e.g., $u_1$($t$), $u_2$($t$), and $u_3$($t$)) into network nodes that drives the system from an initial state to a target state in a fixed period of time. A state is a $1 \times N$ vector $x_{t}$ whose elements represent the activity of each of $N$ nodes in the network at some time $t$. The calculation of minimum control energy estimates the energy required to transmute one state into another allowing activity to spread solely through known inter-regional links. The greater the minimum control energy the more costly and hard-to-reach that target state is said to be. \textbf{(b)} For a given session, we model the network of channels as a linear time independent system and compute the minimum control energy required to transition between chronologically ordered trial states. A trial state is taken to be the firing rate of each channel during a trial. The topology of the network is defined by the session effective connectivity matrix. \textbf{(c-d)} The average minimum control energy (ACE) calculated across all pairwise trial state transitions of a particular session. ACE dynamics for both monkeys across their respective sessions are shown. Filled boundary areas represent +/- 1 standard deviation.}
\label{fig4}
\end{figure*}

In applying and extending network control theory to understand habit formation, our next step is to use the effective connectivity networks to estimate the energy requirement of neural state transitions. We use the concept of minimum control energy from control theory, which represents the minimum amount of input energy necessary to cause a network to transition from a specific initial state of activity to a specific final state of activity (\hyperref[fig4]{Figure 4A}) \cite{stiso2019white}. Intuitively, the more energy a transition requires, the more difficult it is to reach the final state. 

In prior work, minimum control energy has been defined for mechanical and technological systems, or abstract mathematical models. To use the approach here, we must first identify a relevant dynamical model for the considered network process. This model consists of (i) a network state, which we define as the trial firing rates (\hyperref[fig4]{Figure 4B}), (ii) a transition map for the state, which we define as the inferred effective connectivity matrix, and (iii) a set of driver nodes, which include all the nodes in our study (see \hyperref[subsec:methods:controlenergy]{Methods}). With these variables defined, we then estimate the energy required to transmute one state into another allowing activity to spread solely through effective connections. Specifically, we calculated the average minimum control energy (ACE) theoretically required for the observed trial-to-trial state transitions within each session (see \hyperref[subsubsec:methods:ace]{Methods}). The ACE dynamics of both monkeys follow a similar downward trend throughout the entire experiment (\hyperref[fig4]{Figure 4C, 4D}). A simple linear regression confirmed that there was a statistically-significant negative effect between average minimum control energy and session (Monkey G: $\beta$= -.1141 (-.1703, -.0578),  R$^2$= .30, p < $10^{-3}$; Monkey Y: $\beta$= -.0157 (-.0235, -.0080), R$^2$=.13, p < $10^{-3}$). See \textbf{Supplementary Figure 6} for robustness of ACE estimates to variation in model parameters.

\subsubsection*{Relating the Control Energy to Saccades}

\begin{figure*}[!tp]
\includegraphics[width=\textwidth]{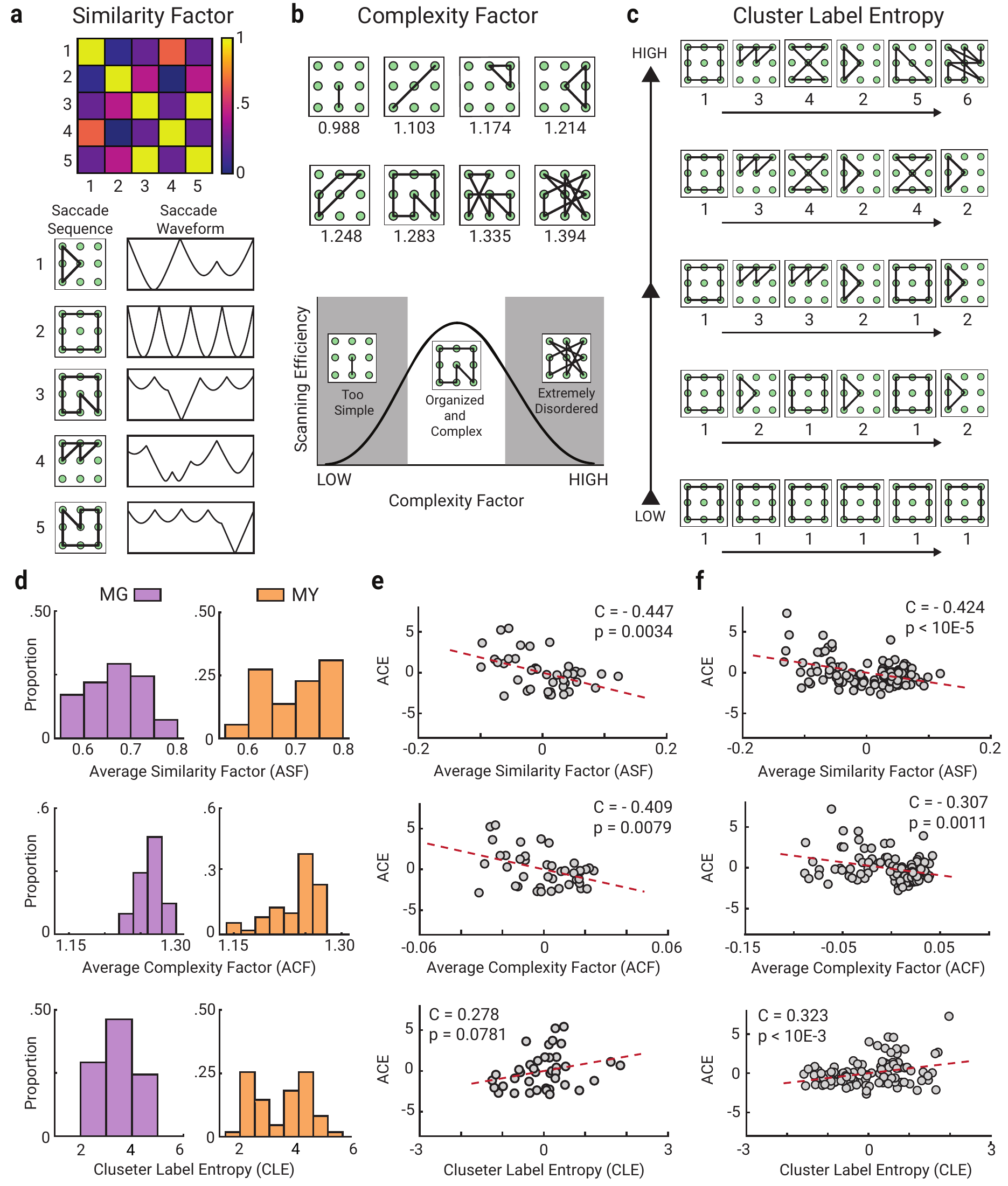}
\caption{\textbf{Relating Control Energy to Saccades}.  \textbf{(a)} The similarity factor (SF) captures information about the similarity between chronologically ordered trial saccade patterns. The similarity between two trial representative saccade patterns is calculated as a metric based on the Euclidean distance between two saccade waveforms. Here we show a visual depiction of 5 example patterns performed by the monkeys, their respective saccade waveform representations, and their similarity to one another. \textbf{(b)} The complexity factor (CF) is calculated as the fractal dimension of the saccade pattern. A range of observed patterns and their complexity factors are shown. Both extremely low and extremely high complexity patterns result in poor scanning efficiency during the task. \textbf{(c)} The cluster label entropy (CLE) metric captures information about the monkey's preference towards exploration of various patterns or exploitation of only a select few during a task session. CLE is calculated as the Shannon entropy of the vector of identified trial cluster labels from a single session. Higher values indicate preference for constantly exploring a variety of clusters with minimal exploitation of any single cluster. \textbf{(d)} Side-by-side comparison of saccade metric distributions across all sessions for both monkeys. The average similarity and average complexity factors were calculated on a per-session basis. \textbf{(e-f)} Pearson correlations between all average saccade metrics and the average minimum control energy (ACE) for Monkey G \textbf{(e)} and Monkey Y \textbf{(f)}. Red-dashed lines are present for visual aide.}
\label{fig5}
\end{figure*}

We next sought to quantitatively characterize how the monkeys' approaches to the free-scanning task changed over time. For this purpose, we defined three metrics: the similarity factor, the complexity factor, and the cluster label entropy. We will discuss each in turn.

\noindent \emph{Similarity Factor.}  We refer to the first metric as the similarity factor (SF), which represents the similarity between two trial representative saccade patterns performed one after another during the same session (see \hyperref[subsubsec:methods:asf]{Methods}). This metric can be used to answer the question, ``Is the monkey performing increasingly similar patterns the longer she engages in the task?''. The higher the value of this metric, the more similar the saccade patterns between trials. It is important to note that this metric was designed to be orientation-independent and reflection-independent. Accordingly, the similarity factor renders two instances of the same pattern as identical even if one was rotated, the patterns were exact mirror images of each other, or rotations of exact mirror images (see \textbf{Supplementary Figure 4}). This feature of the similarity metric is shown in (\hyperref[fig5]{Figure 5A}), where patterns 3 and 5 only differ in their orientation but result in a similarity factor of approximately 1. See \textbf{Supplementary Figure 5a} for the average similarity factor as a function of session for both monkeys.

Although the range of similarity values across task sessions for both monkeys was the same (0.55 to 0.80), the average similarity factor distribution of Monkey Y is skewed towards higher values (\hyperref[fig5]{Figure 5D}). A two-way Kolmogorov-Smirnov test confirmed that the average similarity factor distributions of the two monkeys were significantly different from each other ($D = 0.3541$, $p = 7.57 \times 10^{-4}$). Furthermore, the Pearson correlation between the average similarity factor and average minimum control energy (for Monkey G, $C = -0.447$, $p = 3.4 \times 10^{-3}$; for Monkey Y, $C = -0.424$, $p < 10^{-5}$) was significantly negative in both monkeys (\hyperref[fig5]{Figure 5E, 5F}). Permutation tests were performed independently for each monkey, to ensure that the observed associations between average similarity factor and average minimum control energy were due to the observed neural circuit architecture (see \hyperref[subsubsec:methods:acecorr]{Methods}). The observed correlations between average similarity factor and average minimum control energy for both monkeys proved to be significantly more negative than expected in their respective permutation null distributions (for Monkey G, $p < 10^{-3}$; for Monkey Y, $p < 10^{-3}$).

\noindent \emph{Complexity Factor.} We will refer to the second metric that we defined as the complexity factor (CF), which is a quantitative measure of the complexity of an individual trial representative saccade pattern in a given session. We define complexity as the fractal dimension of the identified pattern (see \hyperref[subsubsec:methods:acf]{Methods}). The metric can be used to answer the question, ``Is the monkey approaching the task in a strategic way or is it simply saccading at random?''. Both extremely low complexity values and extremely high values are not optimal strategies for scanning the task grid efficiently. A pattern with a low complexity ($\approx$1) is often too simple and does not cover all the targets in the task. In contrast, a saccade pattern of high complexity (i.e. greater than 1.3) is extremely disordered, tortuous, and seemingly random without any strategy (\hyperref[fig5]{Figure 5B}). Patterns that strike a balance between organization and complexity offer the most efficient approach to scanning the 3$\times$3 target grid. See \textbf{Supplementary Figure 5b} for the average complexity factor as a function of session for both monkeys.

Both monkeys performed saccade patterns of similar complexity throughout their respective trials, with most sessions averaging to values between 1.24 and 1.28 (\hyperref[fig5]{Figure 5D}). However, the full range of complexity that Monkey Y exhibited was larger than that exhibited by Monkey G, as MY spent several sessions performing markedly simple patterns. A two-way Kolmogorov-Smirnov test confirmed that the average complexity factor distributions of the two monkeys were significantly different from each other ($D = 0.4581$, $p = 3.75 \times 10^{-6}$). Furthermore, the Pearson correlation between the average complexity factor and minimum control energy (for Monkey G, $C = -0.409$, $p = 7.9 \times 10^{-3}$; for Monkey Y, $C = -0.307$, $p = 1.1 \times 10^{-3}$) was significantly negative in both monkeys (\hyperref[fig5]{Figure 5E, 5F}). Permutation tests were performed independently for each monkey, to ensure that the observed associations between the average complexity factor and average minimum control energy were due to the observed neural circuit architecture (see \hyperref[subsubsec:methods:acecorr]{Methods}). The observed correlations between the average complexity factor and average minimum control energy for both monkeys proved to be significantly more negative than expected in their respective permutation null distributions (for Monkey G, $p < 10^{-3}$; for Monkey Y, $p < 10^{-3}$).

\noindent \emph{Cluster Label Entropy.} We will refer to the third metric that we defined as the cluster label entropy (CLE), which is a quantitative estimate of a monkey's preference towards pattern exploration or exploitation during a task session. It is a direct calculation of Shannon's information entropy of the vector of chronologically ordered trial cluster labels in an individual session. The higher the cluster label entropy of a session, the less ordered the behavior and the more prone the monkey was to explore a variety of different saccade patterns coming from multiple identified clusters. The metric can be used to answer the question, ``Is the monkey choosing to explore many different saccade patterns across trials or does it continuously exploit a select few?''. See \textbf{Supplementary Figure 5c} for the saccade cluster entropy as a function of session for both monkeys.

The distribution of cluster label entropy for Monkey Y across sessions shows that she exhibited moments of both extreme exploitation (CLE = 2-3) and extreme exploration (CLE = 4-5.5). In contrast, Monkey G exhibited a preference for mid-range values of cluster label entropy with a majority of the task sessions falling in the range of 3-3.5, a balance between exploitation and exploration (\hyperref[fig5]{Figure 5D}). A two-way Kolmogorov-Smirnov test confirmed that the cluster label entropy distributions of the two monkeys were not significantly different from each other ($D = 0.2020$, $p = 0.1539$).
The Pearson correlation between the cluster label entropy and minimum control energy (for Monkey G, $C = 0.278$, $p = 0.0781$; for Monkey Y, $C = 0.323$, $p < 10^{-3}$) was significantly positive in Monkey Y only (\hyperref[fig5]{Figure 5E, 5F}). Permutation tests were performed independently for each monkey, to ensure that the observed associations between cluster entropy and average control energy were due to the observed neural circuit architecture (see \hyperref[subsubsec:methods:acecorr]{Methods}). The significant correlation between cluster label entropy and average minimum control energy found in Monkey Y also proved to be significantly more positive than expected from the respective permutation null distribution (for Monkey G, $p = 1$; for Monkey Y $p < 10^{-3}$).

\begin{figure*}[!tp]
\includegraphics[width=\textwidth]{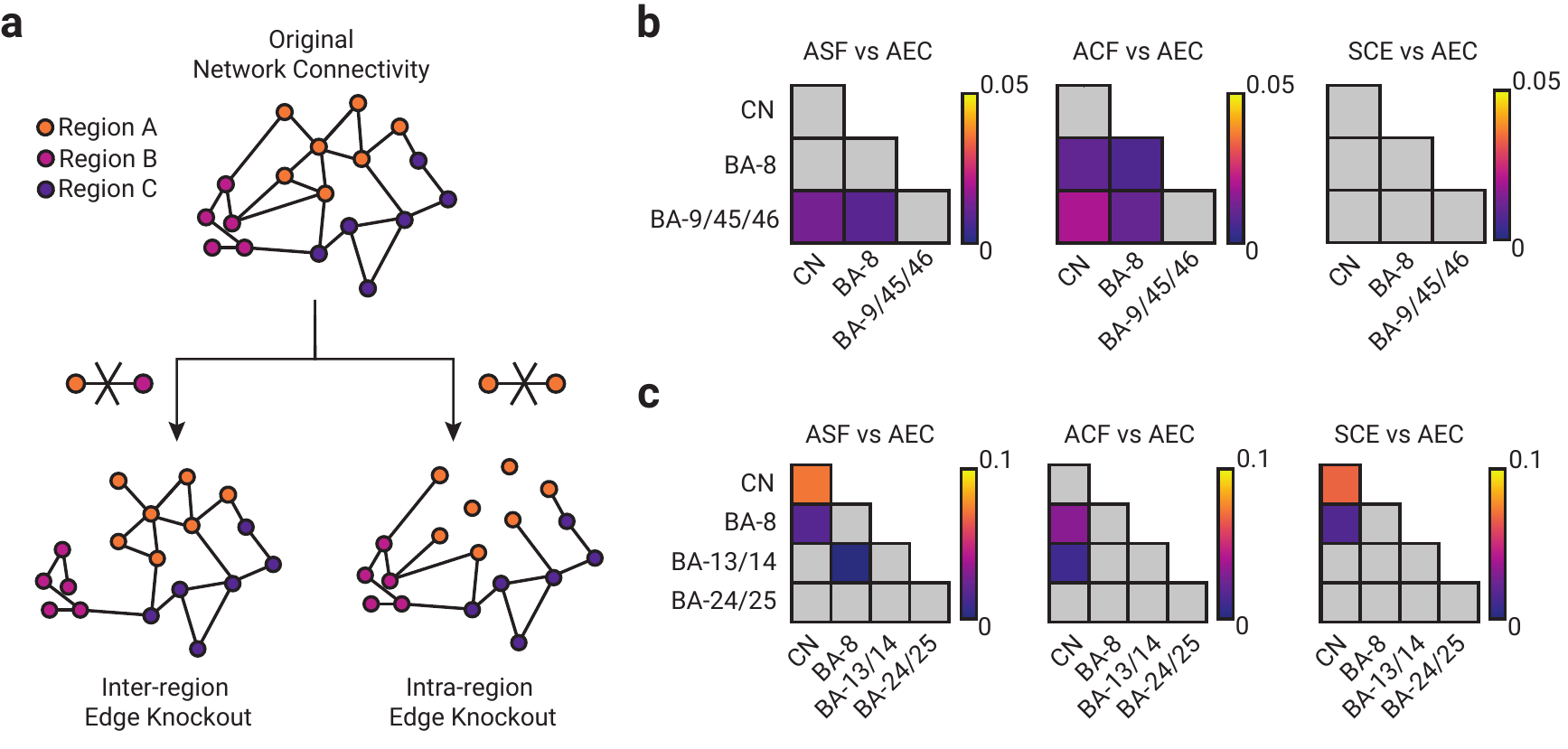}
\caption{\textbf{Virtual Region Specific Lesion Analysis.} \textbf{(a)} Visual depiction of lesion analysis workflow. The lesion knockout consists of performing a series of edge-knockouts where (i) all edges between two regions are set to zero in the effective connectivity matrix (inter-region), or where (ii) all edges connecting one region to itself are set to zero (intra-region). The average minimum control energy is then calculated using the lesioned effective connectivity matrix, and correlations are computed between the average minimum control energy and all saccade metrics. \textbf{(b-c)} Results of lesion analysis across all possible region combinations for Monkey G \textbf{(b)} and Monkey Y \textbf{(c)}. All post-lesion correlations were compared to an equivalent null distribution constructed from random edge lesions. Matrix elements represent the absolute value difference between post-lesion and pre-lesion correlations. All grayed out edges represent lesions which did not result in a significant change in correlation value when compared to their respective null distribution.}
\label{fig6}
\end{figure*}

\subsubsection*{Identifying Neural Substrates Particularly Key to the Relation Between Control Energy and Saccades}

In a final step, we seek to determine which part(s) of the inferred network of brain regions significantly contribute to the observed relationships between control energy and behavior. We do so by performing a virtual lesion analysis consisting of a series of inter- or intra-region edge knockouts in the inferred effective connectivity matrix (see \hyperref[fig6]{Figure 6A} and \textbf{Supplementary Figure 7} for a schematic depiction of this approach). An edge knockout refers to setting the weights of edges in the effective connectivity matrix to a value of zero, thereby virtually removing connections in the network. Edges whose knockout serves to remove the correlation (resulting in a $p$-value greater than $\alpha=0.05$) between average minimum control energy and saccade metrics are inferred to be important in controlling task specific energy dynamics. If the correlations can be removed by localized edge deletions, then we would infer that the energetic constraints on neural state transitions are localized to a particular part of the circuit. If instead the correlations cannot be removed by localized edge deletions, then we would infer that the energetic constraints on neural state transitions are broadly distributed across the circuit.

In a first step, we lesion connections in a manner that is guided by the anatomy, before broadening to an exploratory assessment of random edge lesions. The results from the former are shown in (\hyperref[fig6]{Figure 6B}) and (\hyperref[fig6]{Figure 6C}), respectively. In Monkey G, the removal of connections between BA-8 and BA-9/45/46 resulted in small but significant changes to the originally observed correlation between the control energy and the average similarity factor as well as the average complexity factor. In Monkey Y, the removal of connections within the caudate nucleus resulted in small but significant changes to the originally observed correlation between the control energy and both the average similarity factor and cluster label entropy. Although significant changes were found in both monkeys, none were drastic enough to fully disrupt the observed correlations between behavior and control energy. The observed behavior-energy correlations exhibited resilience to disruptions in the inferred effective connectivity (See \textbf{Supplementary Figure 8}). In order to successfully disrupt the observed correlations in Monkey G, a minimum of 84\% of all edges from its connectivity matrix have to be removed. This minimum threshold increases for Monkey Y, where at least 95\% of all edges have to be set to zero in order to significantly disrupt the correlations. These findings suggest that the energetic constraints on neural state transitions relevant for behavior are only partially localized (\hyperref[fig6]{Figure 6B} \& \hyperref[fig6]{Figure 6C}), but may be more accurately described as being broadly distributed across the circuit.

\section*{Discussion}

Learning commonly requires the development of strategies to increase reward in the face of uncertainty \cite{gottlieb2018towards}. Such strategies can manifest in sequential behaviors that serve to continuously gather information about the environment \cite{desrochers2010optimal}. Yet precisely what rules guide the formation of sequential behaviors remains poorly understood. Although recent work highlights the relevance of distributed circuitry \cite{smith2013dual}, progress has been hampered by the lack of a formal theory linking activity in such circuitry to habitual (or non-habitual) behavior. Here we address this challenge by positing a network control theory of how sequences of behaviors arise from the energy requirements of sequences of neural states occurring atop a complex network structure. Combining behavioral measurements and neural recordings from two female macaque monkeys performing a free-view scanning task over 60--180 sessions \cite{desrochers2015habit,desrochers2010optimal}, we find that our theory predicts smaller energy requirements for transitions between trials in sessions with a high-degree of similarity between complex saccade patterns, and in sessions characterized by an emphasis on the repetition of a small subset of patterns rather than exploration of a more diverse set of distinct patterns. Moreover, we employ a virtual lesioning approach to demonstrate that the derived relationships between control energy and behavior are highly resilient to small, local disruptions in the network, suggesting these observations are associated with the network as a whole rather then a small subset of its nodes. Our study advances a theoretically principled approach to the study of habit formation, provides empirical support for those theoretical principles, and offers a blueprint for future studies seeking to explain how behavior arises from changing patterns of activity in distributed neural circuitry. 

\noindent \textbf{Neural circuits as networks.} The study of habit learning, like the study of many other cognitive functions, has benefited immensely from lesion studies \cite{teng2000contrasting,price2003lesion,izquierdo2004lesion,jenrette2019lesions}, and from focused recordings in single brain regions \cite{yassin2010neocortex,yanike2014representation,kim2015dopamine,desrochers2015habit}. Yet, the field has long appreciated that single regions do not act in isolation, but instead form key components of wider circuits relevant for perception \cite{whitmire2016rapid,garvert2014amygdala}, action \cite{makino2016circuit}, and reward \cite{cox2019striatal}, among others. With recent concerted funding support \cite{litvina2019brain}, many new technologies are now available for large-scale recording of neural ensembles, including methods for high-density multi-region recordings \cite{feingold2012system,chung2019high} and associated novel electrode technologies \cite{hong2019novel}. The advances support a wider goal to gather evermore detailed measurements of activity across the whole brain \cite{kleinfeld2019can}. Here we use multi-channel, multi-area recordings to better understand the distributed nature of neural circuitry underlying habit formation. The channels span Brodmann areas 8 (frontal eye fields), 9 (dorsolateral and medial prefrontal cortex), 13-14 (insula and ventromedial prefrontal cortex), 24-25 (anterior and subgenual cingulate), and 45-46 (\emph{pars triangularis} and middle frontal area), allowing us to probe multi-area activity and inter-areal interactions that track habitual behavior.

Our data naturally motivate the question of how circuit activity supports behavior. This question is certainly not new, and not even specific to neural systems. In fact, the recent rapid expansion of work in artificial neural networks has highlighted the fundamental fact that the architecture of a network is germane to the system's function \cite{veen2019neural}. Liquid state machines \cite{maass2002real}, convolutional neural networks \cite{lecun1998gradient}, and Boltzmann machines \cite{hinton1986learning} all perform distinct tasks defined by their architectures. Similarly, in biological neural systems, empirical and computational evidence links the architecture of projections with the nature of memory retrieval \cite{rajesthupathy2015projections}, flexible memory encoding \cite{curto2012flexible}, sequence learning \cite{rajan2016recurrent}, and visuomotor transformation \cite{murdison2015computations}. Intuition can be drawn from simple small architectures or network motifs \cite{kashtan2005spontaneous} which have markedly distinct computational and control properties \cite{whalen2015observability}. For example, a chain is conducive to sequential processing, whereas a grid is more conducive to parallel processing. Unfortunately, the architecture of multi-area circuits in the primate brain is not quite so simple, thus hampering basic intuitions and straightforward predictions. To address this challenge, we use the mathematical language of network science \cite{mitchell2011complexity}. The network approach allows us to embrace the distributed nature of neural circuit activity and quantitatively describe the empirically observed architecture, while also formalizing questions regarding how that architecture supports circuit function \cite{bassett2018on}.

\noindent \textbf{Activation and effective connection.} Current efforts in computational and systems neuroscience are divided by a focus either on patterns of neural activity or on patterns of neural connectivity. At the small scale, this divide separates studies of the firing rates of neurons from studies of noise correlations \cite{doiron2016mechanics,kohn2016correlations}. At the large scale, this divide separates studies using general linear models or multivoxel pattern analysis in fMRI \cite{friston2005models,mahmoudi2012multivoxel} from studies using graph theoretical or network approaches \cite{bassett2018on}. A key challenge facing the field is the need to span this divide, both in experimental and in theoretical investigations \cite{ocker2017linking}. Indeed, to take the next step in understanding behavior requires the development of computational models of cognitive processes that conceptually or mathematically combine activity and connectivity \cite{bassett2017network}. 

Here we summarize firing rate activity across channels as a brain state, and we probe how such states can change given the effective connectivity between channels. The fact that network architecture can constrain the manner in which activation patterns change (and \emph{vice versa} \cite{schuecker2017fundamental}) is supported by empirical evidence in large-scale human imaging \cite{ito2019discovering}, and a long history of computational modeling studies in human and non-human species \cite{breakspear2017dynamic,ocker2017linking}. Here we inform our modeling choice by noting a particular characteristic of that constraint, which exists in the following form: state $x$ is more likely than state $y$ given network $A$. Specifically, we acknowledge that neural units that are densely connected are more likely to share the same activity profile than neural units that are sparsely connected, for example due to being located in a distant area. This phenomenom naturally arises in many dynamical systems (see, for example, \cite{fan2019enhancing,sorrentino2016complete,TM-GB-DSB-FP:18}), and its study has recently been further formalized in the emerging field of graph signal processing \cite{ortega2018graph,huang2018graph}, which offers quantitative measures to evaluate the statistical relations between a pattern of activity and an underlying graph. 

\noindent \textbf{Network control theory.} Beyond positing a probabilistic relationship between activity and connectivity, we define an \emph{energetic} relation between them. Our formalization utilizes the nascent field of network control theory \cite{motter2015network,pasquletti2014controllability,liu2011controllability}, which develops associated theory, statistical metrics, and computational models for the control of networked systems, and then seeks to validate them empirically. The broad utility of the approach is nicely exemplified in recent efforts that address such disparate questions as how to control the spread of infectious disease in sub-Saharan Africa \cite{roy2011network}, of current in power grids \cite{barany2004nonlinear}, or of pathology in Alzheimer's disease \cite{henderson2019spread}. In the context of neural systems, the theory requires 3 components: (i) a definition of the system's state, such as population-level activity in large-scale cortical areas \cite{stiso2019white} or cellular activity in microscale circuits \cite{yan2017network}, (ii) a measurement of the connections between system components, such as white matter tracts \cite{bernhardt2019temporal} or synapses \cite{towlson2018caenorhabditis}, and (iii) a form for the dynamics of state changes given a network, such as full nonlinear forms \cite{schiff2010neural} or linearization around the current operating point \cite{jeganathan2018fronto}. Here we let the state reflect the firing rate activity across channels, the network reflect effective connectivity between channels, and the dynamics take the form of a linearization around the current operating point. 

After formalizing the theory for a given system, one can use longstanding analytical results to estimate the energy required to move the system from one state to another \cite{hespanha2018linear,boltyanskii1960theory}. We focus specifically on the problem of identifying the minimum control energy, which is a common subform of the more general problem of identifying the control energy required in the optimal trajectory between state $i$ and state $j$ \cite{kailath1980linear}. Outside of neuroscience, the approach has proven useful, for example, in increasing energy efficiency in induction machines \cite{plathottam2015transient}, enhancing performance of transient manufacturing processes \cite{sahlodin2017efficient}, and managing energy usage in electric vehicles \cite{boehme2013trip}, among others. In the context of neural systems, the study of such trajectories has been used to address questions of how the brain moves from its resting or spontaneous state to states of task-relevant or evoked activity, how the brain's network architecture determines which sets of states require little energy to reach, and how electrical stimulation induces changes in brain state \cite{stiso2019white}. Here, we use the approach to estimate the amount of energy that is theoretically needed to push the circuit from the state reflecting firing rate activity in one trial to the state reflecting firing rate activity in the next trial. By performing the calculation for all pairs of temporally adjacent trials, we are able to examine changes in energy over the course of learning. More importantly, structuring the investigation in this way allows us to determine how such energy relates to pairwise differences in sequential behaviors during habit formation.

\noindent \textbf{Energetics of habit formation.} In an expansion upon prior work in the application of network control theory to neural systems, we posit that low energy state transitions characterize processes (and their associated behaviors) that are less cognitively demanding. Informally, the underlying notion harks back across at least two centuries in the history of neuroscience \cite{sourkes2006on}. More formally, we can draw on the theory of maximum entropy \cite{ortega2013thermodynamics}, to posit that transitions between low entropy saccade patterns require greater effort and therefore energy than transitions between high entropy saccade patterns. Note that a high entropy saccade pattern is one that spans many targets in a disordered manner while a low entropy saccade pattern is one that spans few targets in an organized, structured pattern. Concretely, we operationalize the entropy of a trial representative saccade pattern as its fractal dimension, which we refer to as its complexity. Consistent with our hypothesis, we find that the saccade complexity is negatively correlated with the energy theoretically required to move the neural circuit from the firing rate state of one trial to the firing rate state of the next trial. Broadly, our data join that acquired in other model systems, anatomical locations, and species in providing evidence that maximum entropy models explain key features of neural dynamics \cite{savin2017maximum,granot2013stimulus,meshulam2017collective}. 

Moving beyond the assessment of a saccade pattern's entropy, we next consider the role of habits in modulating the cognitive demands elicited by a task \cite{haith2018cognitiveload,moors2006automaticity}. We hypothesize that transitioning between the same (or similar) saccade patterns will require less cognitive effort and therefore less energy, than transitioning between different (or dissimilar) saccade patterns. Consistent with our hypothesis, we find that transitioning between more similar saccades is associated with smaller predicted energy, and that sessions with a larger number of distinct saccade patterns are associated with greater predicted energy. Collectively, these data comprise a formal link between the energetics of neural circuit transitions and sequential behaviors.

\noindent \textbf{The role of localized vs. distributed computations.} By definition, the theoretically predicted energy is a function of both the neural states and the underlying network of effective connectivity, and therefore reflects contributions from all channels and from all inter-channel relations. Nevertheless, it is still of interest to determine whether some channels, or some inter-channel relations, contribute relatively more or less than others \cite{stiso2019white}. Using a virtual lesioning approach, we found that removal of connections in the caudate nucleus, and connections between BA-8, BA-9/45/46 and BA-13/14, resulted in predicted energies that caused small but significant decreases in magnitude to the correlation with saccade metrics. The critical role of the caudate nucleus in habit formation is consistent with prior lesion studies \cite{teng2000contrasting,price2003lesion,izquierdo2004lesion,jenrette2019lesions} and recording studies \cite{yassin2010neocortex,yanike2014representation,kim2015dopamine,desrochers2015habit}. Moreover, the role of prefrontal cortex is consistent with transcranial magnetic stimulation studies showing its necessity for higher-level sequential behavior \cite{desrochers2015necessity}, and its involvement in uncertainty driven exploration \cite{badre2012rostrolateral}. Here we extend these prior studies by demonstrating the relevance of effective connections in these same areas. Our subsequent random lesioning results further extend our understanding of the neuroanatomical support for these behaviors by suggesting that the energetic constraints on neural state transitions are broadly distributed across the circuit.

Another feature of our findings that we find perhaps particularly striking is their specificity to the two monkeys. Monkey G performed more dissimilar saccade patterns from trial-to-trial, consistent with the goal-directed exploration supported by prefrontal connections identified in our lesioning analysis. In contrast, Monkey Y performed more similar saccade patterns from trial-to-trial during sessions, consistent with lower-level habit formation supported by caudate nucleus identified in our lesioning analysis. While our study is underpowered to formally probe individual differences, these preliminary observations motivate future work examining variation in energy-behavior relations in healthy cohorts and disease models.

\subsection*{Methodological Considerations}

Several methodological considerations are particularly pertinent to this work, and here we mention the three that are most salient. First, we note that in understanding the manner in which neural units communicate, one might wish to have full knowledge of the structural wiring between those neural units \cite{schafer2018worm,eichler2017complete}. Despite recent advances in technology at the cellular scale \cite{hillman2019light,eberle2015mission,berger2018vast}, such information is challenging to acquire \emph{in vivo} in large animals, and currently not possible at all in primates. A reasonable alternative is to use the empirical measurements of activity to \emph{infer} the effective relationships between neural units \cite{friston2011functional,tavoni2016neural,schiefer2018from}. Here we take precisely this tack, thereby distilling a weighted connectivity matrix summarizing the degree to which each channel statistically affects another. A marked benefit of effective over structural connectivity between large-scale brain areas is that only the former can be used to study temporal variation on the time scale at which learning occurs \cite{battaglia2012dynamic}. 

A second important consideration pertinent to our work is that we utilize analytical results from the study of linear systems \cite{kailath1980linear} to inform our network control theory approach \cite{pasquletti2014controllability,stiso2019white}. It is well known that neural dynamics -- measured in distinct species and across several imaging modalities -- are in fact nonlinear \cite{breakspear2017dynamic}. Linear models of nonlinear systems are most useful in predicting behavior in the vicinity of the system's current operating point \cite{kim2019linear}, or for explaining coarse time-scale population-average activity (e.g., see \cite{honey2009predicting}). For the study of other sorts of behavior or signals, future work could consider extending our simulations to include appropriate nonlinearities \cite{motter2015network}.

A third important consideration pertinent to our work is the fact that animal behaviors in general -- and saccades in particular -- are complex and difficult to describe cleanly \cite{leshner2011quantification,bellet2019human}. Here we address this difficulty by developing a novel algorithmic approach to the extraction of representative saccade patterns. Our method capitalizes on a graphical representation of saccades, which in turn allows us to use previously developed tools for the characterization of graphs \cite{mitchell2011complexity}. Our effort follows a growing literature using network models to parsimoniously represent and study animal behavior \cite{belyi2017global,hawkins2019emergence}. Publicly available MATLAB code implementing the algorithm may prove useful in the context of similar data, and can be found here https://github.com/kpszym/SaccadePatternExtraction.git. 

\subsection*{Conclusion} Systematically canvassing uncertain environments for reward induces habitual behaviors and engages distributed neural circuits. Here we offer a formal theory based on the principles of network control to account for how pairwise differences in sequential behaviors during habit formation can be explained by the energetic requirements of the accompanying neural state transitions. In doing so, the study frames the concept of cognitive computations within a formal theory of network energetics. Our findings further support the notion that free energy or maximum entropy are useful explanatory principles of behavior. While outside the scope of this study, many relevant questions remain unasked. Future work could usefully expand upon our observations by increasing the number of recorded areas or altering the task to include different sorts of environmental uncertainties. Incorporation of additional computational capabilities into the theory, and exercising agent-based simulations to determine optimal cost functions and associated learning rules for artificial neural systems placed in similar environments could also be used to expand upon this work.

\section*{Methods}
\label{sec:methods}

The data consists of behavioral measurements and neural recordings from two female macaque monkeys: Monkey G (MG) and Monkey Y (MY). Full descriptions are provided in Refs. \citep{desrochers2010optimal,desrochers2015habit}, and here we briefly summarize. Both monkeys were individually monitored, and data was recorded while the monkeys performed a free-viewing scan task. The task was performed across multiple days (sessions) with each session consisting of multiple back-to-back trials. Eye movements were recorded utilizing an infrared tracking system and converted into a sequence of saccades, or rapid eye movements from one point to another. All measurements of neural activity were obtained from individual chronically implanted electrode arrays recording bilaterally from various points in the caudate nucleus (CN), frontal eye fields (FEF) and prefrontal cortex (PFC). \par
    
\subsection*{Task Structure} 

The task begins when a grid of small gray circles is presented on a screen in front of the monkey, whose head is fixed in place. After a variable period of time, the inner gray circles of the grid are replaced with a 2$\times$2 grid or a 3$\times$3 grid of larger green dots (\emph{Targets On}). The monkey’s gaze may not leave the space defined by the perimeter of the green dots or the trial will be marked as unsuccessful and the screen will revert back to a grid of only gray dots. After a variable time, one of the green targets is baited (\emph{Target Baited}) such that if the monkey’s gaze falls on to the baited target, the trial is rewarded. The monkeys were not given information about when the target was baited or which target was baited. At this point in the task, if the monkey’s gaze crossed into or over the bait target, the grid of green dots was replaced by the original gray dots (\emph{Targets Off}). After a variable time, the monkey was presented with a short reward to indicate success; acknowledging their preferences, juice was offered to Monkey G, and an alternative reward mixture was offered to Monkey Y. \par

\subsection*{Data Quality Assurance and Cleaning}

Due to the inherent complexity of the task and behavioral responses thereto, it is critical to apply data quality standards that ensure statistical analyses are appropriate and well-powered.  Accordingly, all analysis involved in inferring effective connectivity was limited to task trials where the monkeys were presented with the 3$\times$3 grid version of the free-scanning task. This criteria resulted in 18,298 available trials for Monkey G and 157,729 for Monkey Y. Analytical steps focused on defining the relationship between task behavior and control energy dynamics were limited to only 3$\times$3 grid task trials that were rewarded and exhibited a looping saccade sequence, which is defined as a sequence that starts and ends at the same grid target. A total of 9,702 (~53\%) task trials were used for Monkey G and 80,664 (~51\%) for Monkey Y. In this particular task, a looping saccade sequence in which all targets were visited once before returning to the starting node is considered optimal \cite{desrochers2010optimal}. Furthermore, the number of available channels, defined as those with non-zero signal, varied across sessions. The 60 recorded sessions for Monkey G contained anywhere from 16 to 38 channels (with an average of 23) from a total of 72 unique channels. The 180 recorded sessions for Monkey Y contained anywhere from 3 to 23 channels (with an average of 11 channels) from a total of 96 unique channels. To ensure adequate sampling, all analysis performed on Monkey Y was limited to sessions containing 8 or more channels. \par

\subsection*{Classification of Saccade Sequences}

\subsubsection*{Conversion to a Saccade Network}

Measurements of monkey eye-movements during the free-scanning task comprise a list of saccades performed in the scanning window of a given trial. Each saccade is represented as a vector of two numbers, which denote the start and end grid targets of the saccade. The entire sequence of saccades performed during a given trial can be written as an $m \times 2$ matrix, $SL$. Each row in $SL$ is a single saccade, with the first and second column representing the start and end targets, respectively. This representation can be thought of as a list of directed connections between points. It is then possible to convert a trial specific sequence of saccades into a directed and weighted graph, which we will refer to as the \emph{saccade network}. 

In graph theory \cite{mitchell2011complexity}, a generic network is made up of $N$ nodes that are connected pairwise by $E$ edges. Here, a saccade network consists of nine nodes ($N = 9$), one for each grid target, and edges defined by the saccade sequence. Specifically, an edge between two nodes in a saccade network exists if a saccade was performed between those two targets. The weight of the edge is given by the number of times that specific saccade was performed. Therefore, the saccade network can be written as the directed and weighted $N \times N$ adjacency matrix, $\mathbf{S}$, whose element $S_{ij}$ denotes the weight of the edge between node $i$ and node $j$. Edges with zero weight signify no connection. 

\subsubsection*{Identifying Trial Representative Saccade Sequences}

We seek to identify unique saccade patterns across trials, which will in turn allow us to investigate how performance strategies might evolve throughout the task \cite{desrochers2010optimal}. We refer to the unique saccade pattern performed during a given trial as the trial representative saccade pattern (TRSP). To identify the TRSP, we utilize the saccade network of a given trial and identify all the cycles in the network. In graph theory, a cycle is defined as a series of edges that allows for a node to be reachable from itself. Therefore, a cycle in the saccade network is a loop that starts and ends on the same target. This cycle can be represented as a binary \textit{cycle matrix}, $\mathbf{L}$, of size $N \times N$; a given element of $\mathbf{L}$ is set to one if the corresponding edge is a part of the cycle. We take the dot product of the cycle matrix $\mathbf{L}$ and the saccade network $\mathbf{S}$, and refer to the resulting matrix as $\mathbf{L}'$. For a given saccade network, multiple cycles can exist, and therefore also multiple $\mathbf{L}'$s. We define the trial representative saccade pattern to be the cycle with the greatest elementwise sum of all weights in its $\mathbf{L}'$, which intuitively is the cycle composed of the most common point-to-point saccades. For an intuitive graphical depiction of this process, see \textbf{Supplemental Figure 1}. 

\subsubsection*{Characterizing Similarity between Trial Saccade Sequences} \label{subsubsec:methods:trsp}

Measuring the similarity between two saccade patterns is difficult in part because many statistics have been developed for the comparison of two graphs, but it is often unclear when one statistic is more or less appropriate than another \cite{wills2019metrics}. To circumvent this issue, it is useful to consider the cyclic path between nodes in a single trial representative saccade pattern to be a 2-D polygon drawn on the 3$\times$3 grid of equally sized and spaced circles presented during the task. Every saccade that is a part of the pattern is represented as a straight line between two centers of circles on the grid. Each line can then be discretized into small segments, allowing it to be summarized as the set of 100 (x,y) coordinates of segment centers. In other words, each saccade pattern can be summarized by the set of $n$ points, $P$:
        \begin{equation}
            P=\{\left(\mathrm{x}_i,\mathrm{y}_i\right)\mid\left(\mathrm{x}_i,\mathrm{y}_i\right)\in R^2\}, \mathrm{for}\;i = 1,\ldots n .
        \end{equation}
which finely samples the lines composing the \textit{saccade polygon}.

While a set of points is a simpler representation than a graph, it remains difficult to compare these point sets in a manner that accounts for the original geometry. To address this difficulty, we first calculate the centroid of the saccade polygon, $C$, and then we calculate the Euclidean distance between $C$ and every point in $P$:
        \begin{equation}
             D_i=\sqrt{\left(P\left(\mathrm{x}_i\right)-C\left(\mathrm{x}_i\right)\right)^2+\left(P\left(\mathrm{y}_i\right)-C\left(\mathrm{y}_i\right)\right)^2}
        \end{equation}
Then $D$ is a 1-D \textit{saccade waveform} that parsimoniously represents the saccade polygon while maintaining geometric information. To ensure comparability across polygons, we interpolate each $D$ to $I=600$ points. \par

The fact that the saccade waveform can be thought of as a time series representation of a saccade pattern informs our measure of similarity. Importantly, we wish our measure to be invariant to rotation and reflection, such that two polygons rotated by 90 degrees from one another or two polygons which are direct mirror images of each other are correctly determined to be the same. Therefore, rather than simply calculating the Euclidean distance between two saccade waveforms, we instead calculated the Euclidean distances between one saccade waveform and a series of circularly shifted versions of the second saccade waveform. A circular shift is a mathematical operation where a vector is rearranged such that the last element is moved to the first position and all other elements are shifted forward by one. By performing this operation $l$ times, it is possible to shift the last $l$ values to the front of the vector and all other values forward by $l$ positions. 

Accordingly, we create a set of circularly shifted saccade waveforms that represent rotations of the original saccade sequence by different angles as well as rotations of the mirror image of the original saccade sequence. The mirror image of the saccade sequence can be represented by flipping the saccade waveform from left to right (see \textbf{Supplementary Figure 4}). We write the angle of rotation as $\alpha=360(\frac{l}{I})$. For each pair of saccade sequences, a two-step approach is used to quantify their dissimilarity. First, for each pair we calculated the Euclidean distances between one saccade waveform and the circular shifts of a second saccade waveform (including its mirror image representation) in intervals of $\alpha=6$ degrees such that $l\approx10$. From this set of calculations we identify the circular shift which resulted in the smallest Euclidean distance and denote its index as $i_{min}$. Next, we repeated the above process but now performed shifts of size $l=1$ such that $\alpha\approx0.58$. These fine-grained secondary calculations included only circular shifts between $i_{min}-1$ and $i_{min}+1$. The dissimilarity factor (DF) between the two saccades was taken to be the minimum of the calculated distances in the second step; saccades with large distances between them are more dissimilar.

\subsubsection*{The Average Similarity Factor} \label{subsubsec:methods:asf}

To summarize the saccade pattern similarity between two adjacent trials within a task session, we defined the \textit{similarity factor} (SF). For each session $s$ containing $T_s$ trials, the saccade dissimilarity was calculated between saccade patterns from pairs of consecutive trials as described in the previous section. Each value was then converted into a measure of similarity as follows: $SF = 1 - \frac{DF}{DF_{max}}$, where $DF_{max}$ is the maximum dissimilarity factor observed out of all trials from both monkeys. For each session the \textit{average similarity factor} was then given by the average of all similarity factors calculated for that session. See \textbf{Supplementary Figure 5a} for the average similarity factor as a function of session for both monkeys.

\subsubsection*{The Complexity Factor} \label{subsubsec:methods:acf}

To characterize the complexity of each saccade pattern, we defined the \textit{complexity factor}. We operationalized the notion of complexity as the fractal dimension \cite{mandelbrot1983fractal,iannaccone1996fractal}, which has proven useful in the study of many other biological \cite{smith1996fractal,liu2003fractal} and network systems \cite{song2005self}. For each trial representative saccade pattern, we first constructed a binary image of its polygon representation, where the background was black and the saccade pattern outline was white. We then applied the box-counting method to this image to compute the fractal dimension \cite{li2009improved}. Due to the nature of this method, two identical patterns rotated by 90 degrees from one another would result in different fractal dimension values. Therefore, the final fractal dimension value for each trial pattern was taken to be the minimum calculated from the set of all possible rotations of the pattern and its mirror image at intervals of 90 degrees. For each session, $s$, containing $T_s$ trials, the \textit{average complexity factor} was given by the average fractal dimension of all trial representative saccade patterns in that session.  See \textbf{Supplementary Figure 5b} for the average complexity factor as a function of session for both monkeys.

\subsubsection*{The Cluster Label Entropy} \label{subsubsec:methods:sce}

To quantify the extent to which the monkey selects saccade patterns from trial to trial in an ordered fashion, we defined the \textit{cluster label entropy} metric. More precisely, our goal was to determine whether the monkey was randomly performing various patterns or selectively repeating only a few unique ones. We began by using the MATLAB function \textit{linkage()} to perform agglomerative clustering on the saccade waveforms from all the trials of an individual monkey \cite{rokach2005clustering}. The algorithm outputs a hierarchical, binary cluster tree, also known as a dendrogram, based on an input of a $T \times T$ distance matrix, where the $ij$-th element gives the dissimilarity factor $DF$ between the saccade pattern of trial $i$ and the saccade pattern of trial $j$. The height of a link between two objects in the dendrogram directly denotes the distance between those two objects in the data. \par

It is important to note that the dendrogram itself does not indicate the optimal number of clusters that the data should be split into; rather, it demonstrates the order in which objects should be clustered. However, there is a way to identify the natural divisions of the data into distinct clusters using the derived cluster tree. The inconsistency coefficient metric is used to compare the height of a link in a cluster tree to heights of all the other links underneath it in the tree. If the difference is dramatic, it signifies that that newly formed group consists of linking two highly distinct objects. Imposing a threshold on the inconsistency coefficient during clustering captures more natural divisions in the data rather than arbitrarily setting a maximum number of possible clusters. \par

Accordingly, using the MATLAB function \textit{cluster()} we constructed clusters from the hierarchical cluster tree using a range of inconsistency coefficient values (0.1-1.5 in intervals of 0.05) as a threshold criterion, and then calculated the average within cluster sum-of-squares. Using the elbow-method and the calculated within cluster sum-of-squares, we selected an inconsistency coefficient of 0.95 to be the optimal threshold criterion for clustering. This choice resulted in a total of 136 clusters being identified for Monkey G and 346 for Monkey Y. See \textbf{Supplementary Figures 2 and 3} for a detailed listing of cluster patterns of both monkeys.\par

After coarse-graining the data by identifying saccade clusters across trials and sessions, we next turned to assessing whether the monkey transitioned between saccades of the same cluster or of different clusters, and to what degree. We began by identifying the representative saccade sequence of each discovered cluster by finding the trial representative saccade pattern that had the minimum sum of the dissimilarity factor when compared to all other within-cluster patterns. For each session, we then calculated the \textit{cluster label entropy} as Shannon's information entropy of a given session's vector of trial cluster labels.  See \textbf{Supplementary Figure 5c} for the saccade cluster entropy as a function of session for both monkeys.

\subsection*{Channel Firing Rates} \label{subsec:methods:fr}

Information about neuronal activity was not available for all channels during each session; some channels, which we refer to as non-viable channels, showed no activity across all task trials. All non-viable channels were discarded prior to data analysis. MG contained a total of 59 viable channels and MY contained a total of 64 viable channels. On average, a given free-scanning task session contained 23 active channels for MG and 11 active channels for MY. While biologically expected, this variation in channel availability across sessions can adversely affect estimates of effective connectivity (see next section). For example, if we were to compute the effective connectivity from the activity of all channels across all trials at the same time, we could obtain spurious results due to the incomplete sampling across trials and time. \par

To mitigate potential biases due to variable channel availability, we estimate the effective connectivity in each session separately. Every session contains $N_{CH}$ channels from which a signal is available. The signal from each channel is in the form of a spike train, or a vector of ones signifying neuronal activation, and the time at which each activation occurred. All spikes from the full duration of a trial were used when inferring effective connectivity. For analysis concerned with identifying the relationship between behavior and control energy, we focused solely on spikes that occurred after the task grid was presented to the monkeys ($t_{targets\;on}$) and before the task grid disappeared signifying success ($t_{targets\;off}$). This window of time is referred to as the scanning period ($t_{sp}$). 

The firing rate $r$ of a single channel is given by the number of spikes per second.  Calculation of the fire rate of all channels during an individual trial then results in a $1 \times N_{CH}$ vector that represents the activation state of the channel network during that trial. We will refer to this column vector as the neural state $x_{t}$:
        \begin{equation}
            x_t=\left[\begin{matrix}r_1\\r_2\\\vdots\\r_{N_{CH}}\\\end{matrix}\right]\mathrm{, ~~where~t=}1,\ldots T_s.
        \end{equation}
State vectors $x_t$ are calculated for each trial, across all sessions, and for each monkey. 
        
\subsection*{Inferring Effective Connectivity} \label{subsec:methods:ec}

Effective connectivity provides information that differs from both functional connectivity and structural connectivity \cite{friston2011functional}. For nearly three decades, effective connectivity has been ``understood as the experiment and time-dependent, simplest possible circuit diagram that would replicate the observed timing relationships between the recorded neurons'' \cite{aertsen1991dynamics}. The pattern of effective connectivity among many units can be usefully represented as a network, composed of nodes (neural units) and edges (effective connections) derived from node activity. Here, we construct such a network for the set of electrode channels used to record neuronal activity during trials of the free-scanning task. We chose transfer entropy as the method to estimate effective connectivity \cite{vicente2011transfer}, although we acknowledge that other methods exist and could similarly prove useful in the study of habit learning. For each session, we computed the transfer entropy between all pairs of viable-channels from the set of $z$-scored state vectors $x_t$ of size $1 \times N_{CH}$ to obtain an $N_{CH} \times N_{CH}$ directed, unsymmetrical effective connectivity matrix whose diagonal elements are set to zero. All calculations of transfer entropy were performed using \textit{calc_te()} function from the \textit{RTransferEntropy} package in R. \par

For completeness, we gather all session effective connectivity matrices into the 3-D matrix $\mathbf{M}$ whose element $M_{i,j,k}$ represents the effective connectivity between channels $j$ and $k$, derived from the $i^{th}$ session. From the individual session effective connectivity matrices, we calculated the overall effective connectivity matrix, $\mathbf{\mathcal{M}}$, whose element $\mathbf{\mathcal{M}}_{i,j}$ is given by the average of the set of the non-zero connection strengths between channels $i$ and $j$ derived from only the sessions in which both channels were available. Accordingly, $\mathbf{\mathcal{M}}$ is a square directed and unsymmetrical matrix of size $N_{TC} \times N_{TC}$, where $N_{TC}$ is the total amount of available channels across all sessions for an individual monkey.

\subsection*{Network Control Theory} \label{subsec:methods:controlenergy}

To build an intuition for how we use network control theory to probe relations between neural circuit activity and behavior, we begin with a few preliminaries. We consider a nonlinear dynamical system and linearize those dynamics about the system's current operating point \cite{kailath1980linear}. The dynamics of the resultant linear time invariant (LTI) system can be written as:
        \begin{equation}
            \dot{x}=\mathbf{A}x\left(t\right)+\mathbf{B}u\left(t\right)
        \end{equation}
where $N$ is the number of nodes, $\mathbf{A}$ is the $N \times N$ adjacency matrix, $x\left(t\right) = \left[x_1\left(t\right),x_2\left(t\right),\ldots x_N\left(t\right)\right]$ is the state of all network nodes at time t, $u\left(t\right)=\left[u_1\left(t\right),u_2\left(t\right),\ldots u_K\left(t\right)\right]$ gives the external control input for $K$ number of \textit{driver nodes} which receive external input in order to drive the state change of the network. In this work, all network nodes are set to be driver nodes ($K = N$) and as such $\mathbf{B}$ is the $N \times N$ identity matrix. 

\subsubsection*{Average Minimum Control Energy} \label{subsubsec:methods:ace}

The minimum control energy is defined to be the minimum amount of energy that a controller requires to drive an LTI system from some initial state to a target final state in a specified amount of time \cite{kailath1980linear}. There exists an input vector $u\left(t\right)$ that can move the defined network from its initial state, $x_o$, to a state $x_f$ in time $t_f$, with the minimum energy expenditure, $E_{min}(t_f)=\int_{0}^{t_f} \left\lVert u(\tau)\right\rVert ^{2}d\tau$. Practically, we can calculate the minimum control energy to reach the target network state ($x_f$) from an initial state ($x_o$) as
        \begin{equation}
            E_{min}\left(t_f\right)=\left(e^{\textbf{A}t_f}x_o-x_f\right)\textbf{W}_c^{-1}\left(t_f\right)\left(e^{\textbf{A}t_f}x_o-x_f\right),\ 
        \end{equation}
where $T=t_f$ is the time horizon and $W_c^{-1}\left(T\right)$ is the controllability Gramian,
        \begin{equation}
            W_c^{-1}\left(t_f\right)=\int_{0}^{t_f}{e^{\textbf{A}\tau}BB^Te^{\textbf{A}^T\tau}}d\tau
        \end{equation}
for the system. The time horizon is set to a value of 1 for all calculations in this analysis. 

Here we use this framework to compute the average minimum control energy required to move the neural circuit from firing rate state $x_t$ to firing rate state $x_{t+1}$ given the effective connectivity $\mathbf{A}_s = \mathbf{\mathcal{M}}_{i,j}$ for all channels in that session. Note that $x_t$ and $x_{t+1}$ are the firing rate states of two consecutive trials within a given session. Thus, we obtain an $E_{min}$ value for every consecutive trial pair. The \textit{average minimum control energy} (ACE) metric is then the average of $E_{min}$ across all state transitions in that session.

\subsubsection*{Control Energy \& Saccade Characteristics} \label{subsubsec:methods:acecorr}

We calculated Pearson correlation coefficients between the three saccade characteristic metrics (similarity factor, complexity factor, cluster label entropy) and the average minimum control energy. The number of channels (variable across sessions) was regressed out of all variables included in correlations. To ensure that all correlations were specific to the control energy dynamics derived from the inferred overall effective connectivity matrix, $\mathbf{\mathcal{M}}$, we permuted that vector of average control energy (1000 times) and recalculated all Pearson correlations per permutation. A one-tailed test was then used to determine the significance of the Pearson correlation coefficients calculated with the original average control energy dynamics against their respective null distributions, at a significance level of $\alpha=\ 0.05$. 

\subsection*{Network Region Lesion Analysis} \label{subsubsec:methods:lesion}

For both monkeys, electrode channels recorded bilaterally from regions of the caudate nucleus, frontal eye fields, and prefrontal cortex. In order to examine the extent to which specific nodes or edges contribute to the inferred overall effective connectivity matrix, we performed a virtual lesion analysis. Here a lesion is operationalized by setting specific elements $\mathbf{\mathcal{M}}_{i,j}$ of the effective connectivity matrix to zero, thereby effectively eliminating the connection between the $i^{th}$ and $j^{th}$ nodes. Each monkey had a unique set of regions, $R=\{R_1,R_2,\ldots R_N\}$, from which the $N_{TC}$ channels were recording such that each channel was assigned only one region across the entire duration of the task. 

For each monkey, we performed two types of lesions. First, we lesioned all the edges between any two regions, $R_i$ and $R_j$, and we refer to this method as the \textit{inter-region edge knockout}. Second, we lesioned all edges belonging to the same region, and we refer to this method as the \textit{intra-region edge knockout}. Each lesion results in a knockout effective connectivity matrix which we denote as $\mathbf{\mathcal{M}}^{KO}$. Therefore, if performing an inter-region edge knockout between the caudate nucleus ($R_1$) and Brodmann Area 8 ($R_2$) then $\mathbf{\mathcal{M}}^{KO}$ would be the same as the original effective connectivity matrix, $\mathbf{\mathcal{M}}$, except that all elements that represent connections between $R_1$ and $R_2$ are set to zero. In the same way, if performing an intra-region edge knockout lesion of the caudate nucleus then $\mathbf{\mathcal{M}}^{KO}$ would be the same as $\mathbf{\mathcal{M}}$ but have all elements that represent connections of caudate nucleus nodes to other caudate nucleus nodes set to zero. \par

To determine the relevance of a connection for an energy-behavior correlation, we used two criteria. The first criterion was that the lesion resulted in a correlation value that was not significantly different ($p>0.05$) from that obtained using the original permutation null model (random permutations of the original average minimum control energy vector values). The obtained p-value is referred to as $p_{general}$ (\textbf{see Supplementary Figure 7b and 7d}). To assess this criterion, we calculate the knockout ACE metric, $ACE^{KO}$, for each lesion and use it to recompute the Pearson correlation values between $ACE^{KO}$ and the average saccade metrics. We test the significance of each knockout correlation value using a one-tailed test against the null distribution derived from calculations involving the original permutation null model. A knockout correlation that fails to prove significant from the above one-tailed test signifies that the inter- (or intra-) region edges knocked out are important to the inferred effective connectivity matrix and its relationship to task behavior. 

The second criterion for determining the relevance of a connection for an energy-behavior correlation was that the lesion-induced disruption of the observed correlation was specific to the lesion chosen, and not expected by lesioning the same number of randomly chosen edges. To assess this criterion, every knockout correlation value is then compared to the original correlation value by calculating the absolute value of the difference between them, resulting in a correlation difference metric for each lesion. To ensure that the difference in correlations is truly related to the knocking out of the lesion specific edges, the results are tested using the following null model. We state the null hypothesis that for a given lesion consisting of knocking out $n$ specific edges $E^{KO}=\{e_1,e_2,\ldots.e_n\}$, the resulting correlation difference value is no different than the correlation difference value derived from knocking out the set of $n$ randomly selected edges, $E^{null}=\{e_i\mid e_i\notin E^{KO}\}$. Therefore, for every lesion a null distribution of 1000 correlation values was created by randomly knocking out the same number of edges as the original lesion with no one edge being the same as any knocked out in the original lesion. All values were tested against their respective null distribution using a one-tailed test with a significance level of $\alpha=\ 0.05$. The obtained p-value is referred to as $p_{lesion}$ (\textbf{see Supplementary Figure 7c and 7e}). Gray boxes shown in (\hyperref[fig6]{Figure 6B-C}) indicate edges which when lesioned did not result in a significant change in behavior-energy correlation when compared to the respective null distribution derived from randomly lesioning edges.

Due to the small magnitude changes in behavior-energy correlations caused by inter-/intra-region virtual lesioning, we next sought to quantify the number of edge lesions that are required to completely disrupt an observed behavior-energy correlation. Accordingly, for each monkey the resilience of each behavior-control correlation was tested by performing increasingly large lesions and re-calculating the correlation values. The analysis started with 1-edge lesions and ended with lesions involving up to 95\% of all available edges. Each lesion was performed 100 times with random edges and the average changes in behavior-energy correlation are shown in \textbf{Supplementary Figure 8} for Monkey G and Monkey Y.

\section*{Acknowledgments} We thank Jason Z. Kim, Eli J. Cornblath, Pragya Srivastava, Christopher W. Lynn, Harang Ju, Sophia David, Jennifer A. Stiso, William Qian, and David M. Lydon-Staley for helpful comments on earlier version of this manuscript. The work was primarily supported by an ARO MURI awarded to Bassett \& Graybiel (Grafton-W911NF-16-1-0474). The work was further supported by the John D. and Catherine T. MacArthur Foundation, the Alfred P. Sloan Foundation, the ISI Foundation, the Paul Allen Foundation, the Army Research Laboratory (W911NF-10-2-0022), the National Institute of Mental Health (2-R01-DC-009209-11, R01-MH112847, R01-MH107235, R21-M MH-106799), the National Institute of Child Health and Human Development (1R01HD086888-01), National Institute of Neurological Disorders and Stroke (R01 NS099348), and the National Science Foundation (NSF PHY-1554488, BCS-1631550, and IIS-1926757). The content is solely the responsibility of the authors and does not necessarily represent the official views of any of the funding agencies.

\section*{Citation Diversity Statement} Recent work in several fields of science has identified a bias in citation practices such that papers from women and other minorities are under-cited relative to the number of such papers in the field \cite{mitchell_gendered_2013,dion_gendered_2018,caplar_quantitative_2017, maliniak_gender_2013,dworkin2020extent}. Here we sought to proactively consider choosing references that reflect the diversity of the field in thought, form of contribution, gender, and other factors. We obtained predicted gender of the first and last author of each reference by using databases that store the probability of a name being carried by a woman \cite{dworkin2020extent,zhou2020gender}. By this measure (and excluding self-citations to the first and last authors of our current paper), our references contain 10.6\% woman(first)/woman(last), 5.8\% man/woman, 19.2\% woman/man, 57.7\% man/man, and 6.73\% unknown categorization. This method is limited in that a) names, pronouns, and social media profiles used to construct the databases may not, in every case, be indicative of gender identity and b) it cannot account for intersex, non-binary, or transgender people. We look forward to future work that could help us to better understand how to support equitable practices in science.

\subsection*{References}

\bibliography{bibs.bib}

\end{document}

% \maketitle
\thispagestyle{fancy}

\newpage
\setlength{\headsep}{5pt plus 2pt minus 2pt}
\thispagestyle{fancy}
\begin{figure*}[!ht]
\includegraphics[width=\textwidth]{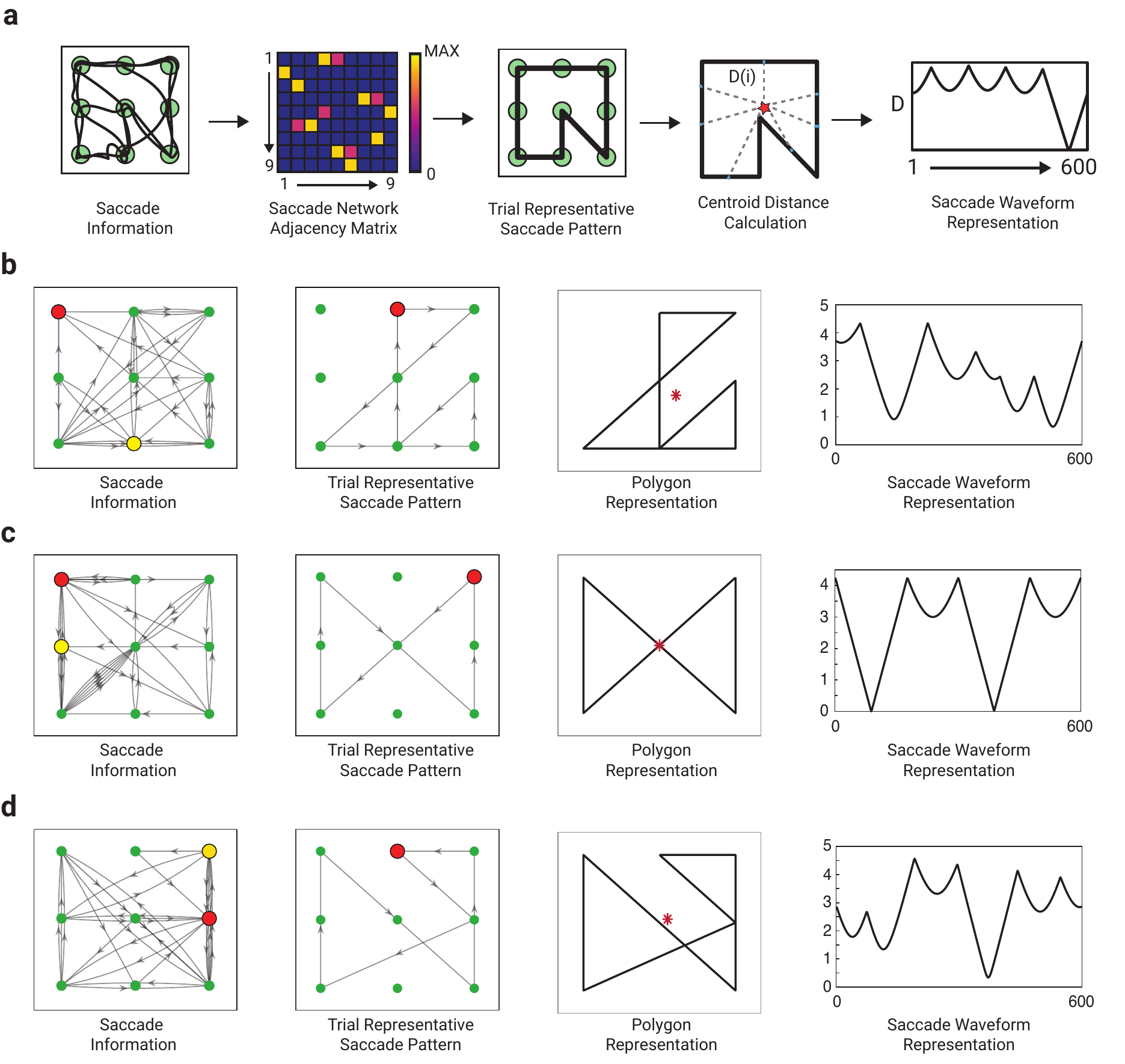}
\captionsetup{labelfont={small, bf}, textfont=small}
\caption{\textbf{Classification of Trial Representative Saccade Patterns}. \textbf{(a)} Saccade information in the form of identified saccadic movements during a trial is collectively represented as an adjacency matrix, which in turn encodes a directed and weighted network. A total of nine nodes exist: one for every green target on the task grid. Edgeweights are calculated as the number of times that a saccade is made from one node to another. The network is converted into a trial representative saccade pattern by identifying the network cycle with the greatest sum of edge weights along its path. Each trial representative saccade pattern is treated as a 2-D polygon in the task grid space consisting of a set of (x,y)points. The saccade waveform is taken to be the vector of Euclidean distances between the polygon centroid and all of its points. A one dimensional interpolation is performed to reduce each saccade waveform to 600 values. \textbf{(b,c,d)} Example step-by-step classifications of saccade patterns from three randomly generated lists of saccades. Yellow targets denote the starting point of the saccade information, while the red target marks the ending point. Since all trial representative saccade patterns are loops, the start and end targets are the same. The red star located on the polygon representations of the saccade patterns marks the centroid of the polygon.}
\label{supp_fig1}
\end{figure*}

\newpage
\thispagestyle{fancy}
\begin{figure*}[!ht]
\includegraphics[width=\textwidth]{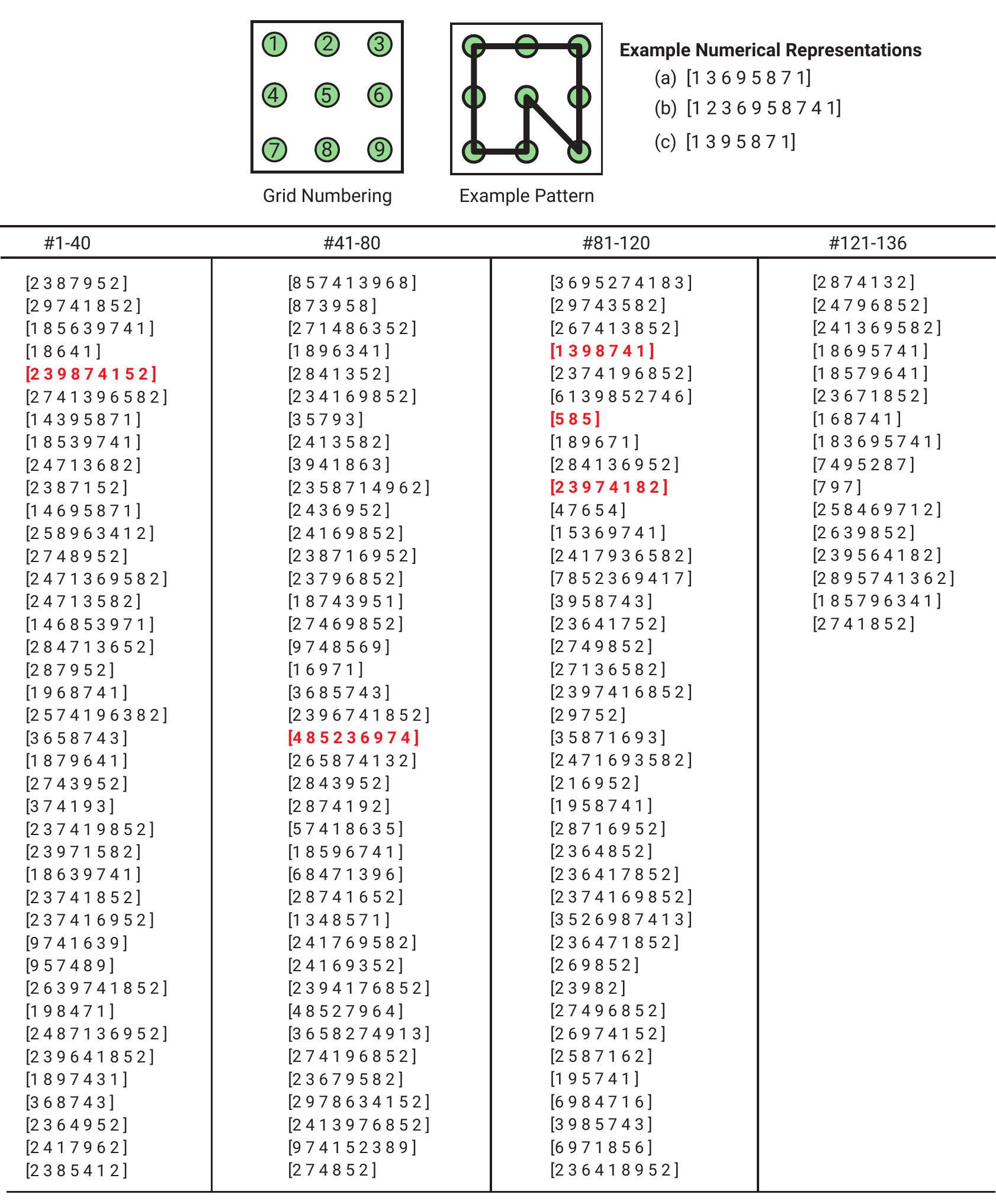}
\captionsetup{labelfont={small, bf}, textfont=small}
\caption{\textbf{Representative Cluster Saccade Patterns for Monkey G}. The 136 identified saccade pattern clusters exhibited by Monkey G are shown in their numerical representations. The diagram at the top demonstrates how a saccade pattern is converted into a numerical representation. Each target on the grid is labeled as a number 1-9. The numerical representation of a pattern then follows to be the numerical sequence of target indices listed in the order that they would be visited when tracing out the pattern. Each cluster numerical sequence identifies the saccade pattern which was most similar to all other patterns in its cluster. The five most prominent clusters are marked by red type.}
\label{supp_fig2}
\end{figure*}

\newpage
\thispagestyle{fancy}
\begin{figure*}[!ht]
\includegraphics[width=\textwidth]{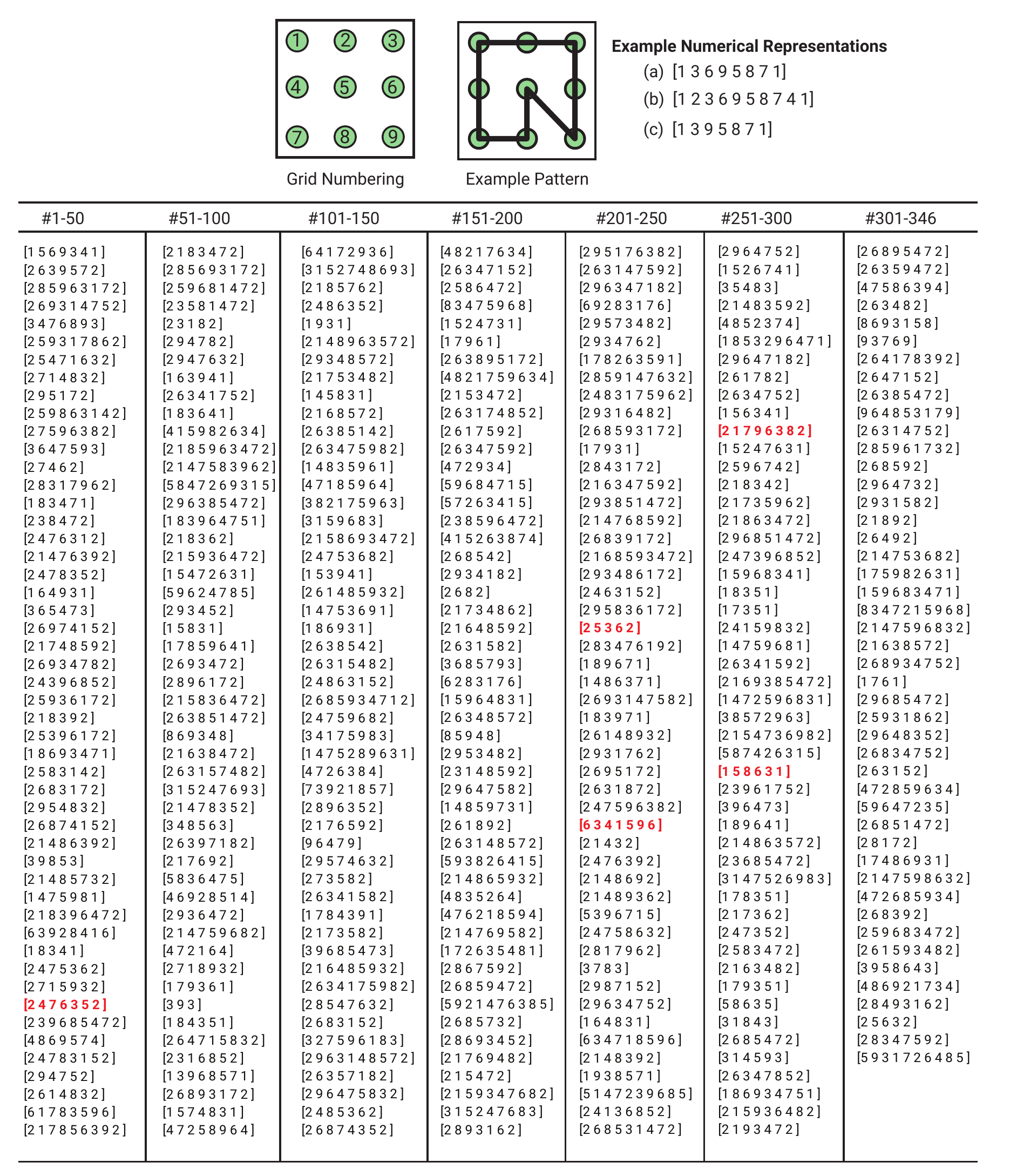}
\captionsetup{labelfont={small, bf}, textfont=small}
\caption{\textbf{Representative Cluster Saccade Patterns for Monkey Y}. The 346 identified saccade pattern clusters exhibited by Monkey Y are shown in their numerical representations. The diagram at top demonstrates how a saccade pattern is converted into a numerical representation. Each target on the grid is labeled as a number 1-9. The numerical representation of a pattern then follows to be the numerical sequence of target indices listed in the order that they would be visited when tracing out the pattern. Each cluster numerical sequence identifies the saccade pattern which was most similar to all other patterns in its cluster. The five most prominent clusters are marked by red type.}
\label{supp_fig3}
\end{figure*}

\newpage
\thispagestyle{fancy}
\begin{figure*}[!t]
\includegraphics[width=\textwidth]{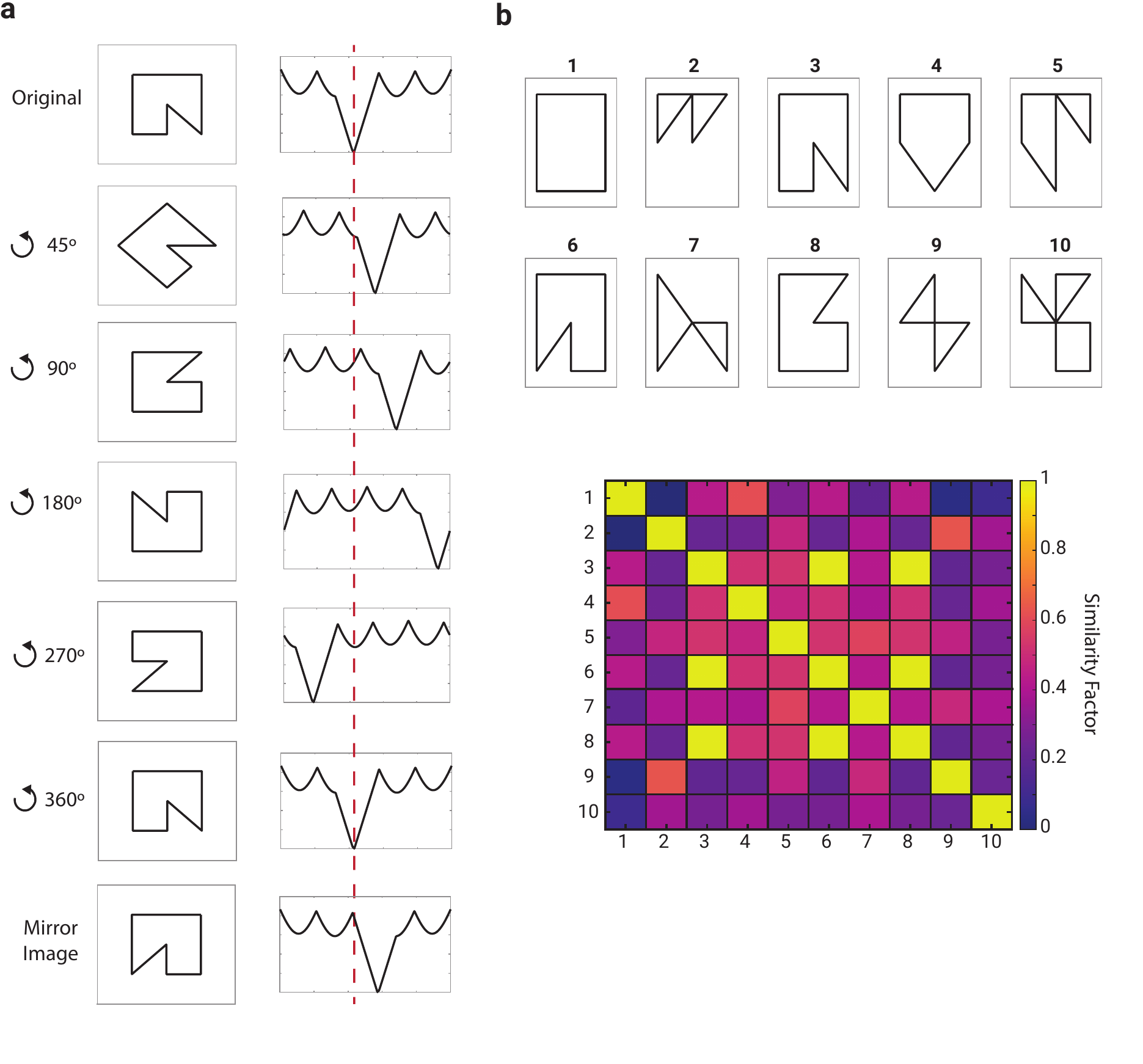}
\captionsetup{labelfont={small, bf}, textfont=small}
\caption{\textbf{Rotational Independence of the Similarity Factor}. The similarity factor metric to compare saccade patterns from trial to trial was designed to be independent of rotation. \textbf{(a)} Performing circular shifts to the saccade waveform is equivalent to rotation of the saccade polygon. A circular shift is a mathematical operation where a vector is rearranged such that the last element is moved to the first position and all other elements are shifted forward by one. By performing this operation $l$ times, it is possible to shift the last $l$ values to the front of the vector and all other values forward by $l$ positions. This relationship is depicted as the pattern rotates counter clockwise, the waveform shifts all elements forward. In addition, the mirror image of the original polygon can be represented by flipping the original saccade waveform left-to-right. The red-dashed line is meant to serve as a visual aid. \textbf{(b)} Ten arbitrary saccade patterns and their calculated similarity matrix. The rotational independence of the measurement is evident as the value of similarity between pattern 8 and patterns 3 and 8 is equal to 1 (the highest value). Note that the value of similarity between pattern 6 and 3 (direct mirror images) is also equal to 1.}
\label{supp_fig4}
\end{figure*}

\newpage
\thispagestyle{fancy}
\begin{figure*}[!t]
\includegraphics[width=\textwidth]{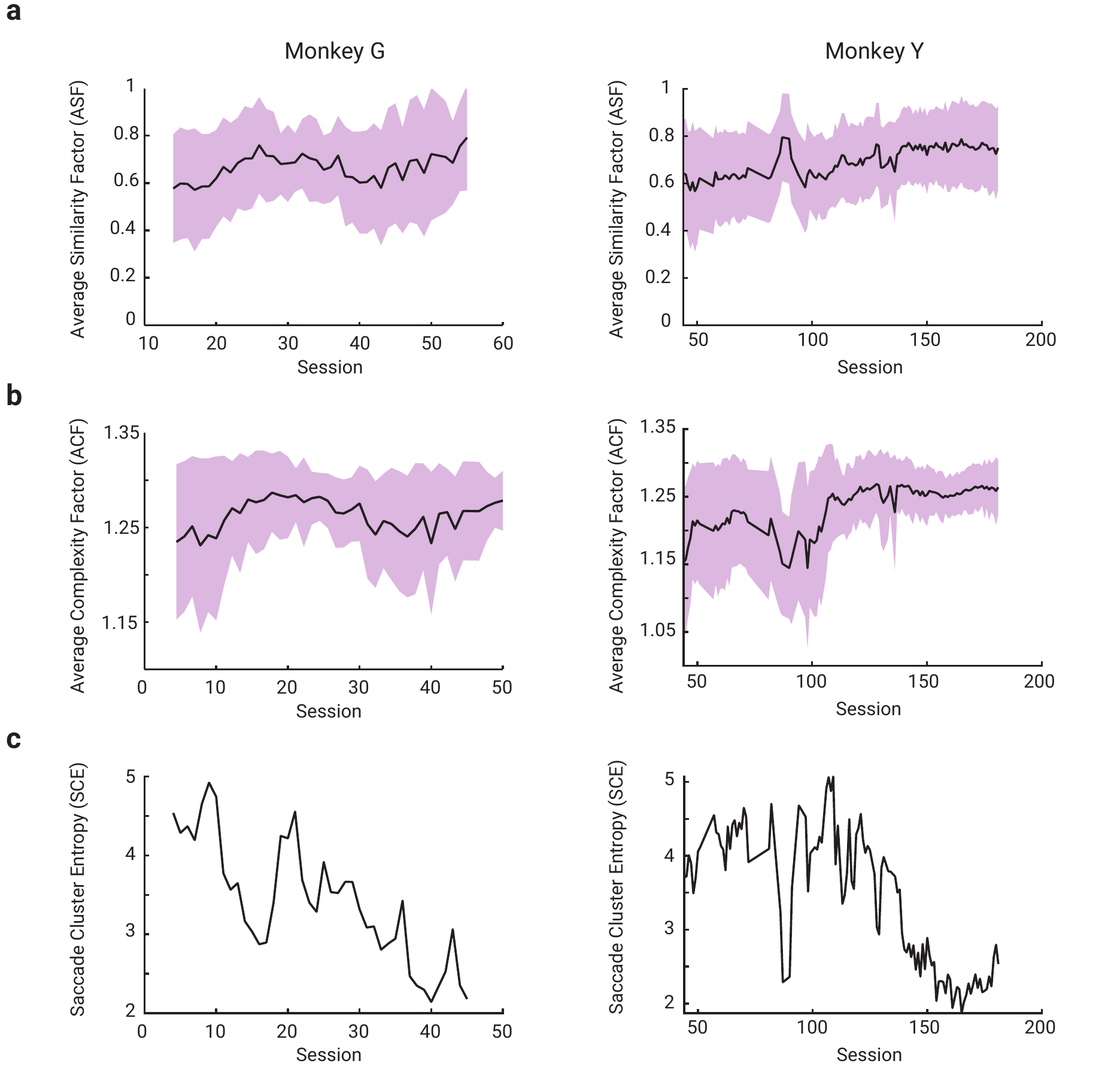}
\captionsetup{labelfont={small, bf}, textfont=small}
\caption{\textbf{Saccade Metric Dynamics}. \textbf{(a)} Dynamics of the average similarity factor across all sessions for Monkey G (Left) and Monkey Y (Right). Filled boundary areas represent +/- 1 standard deviation. \textbf{(b)} Dynamics of the average complexity factor across all sessions for Monkey G (Left) and Monkey Y (Right). Filled boundary areas represent +/- 1 standard deviation. \textbf{(c)} Dynamics of the cluster label entropy across all sessions for Monkey G (Left) and Monkey Y (Right).}
\label{supp_fig5}
\end{figure*}

\newpage
\thispagestyle{fancy}
\begin{figure*}[!t]
\includegraphics[width=\textwidth]{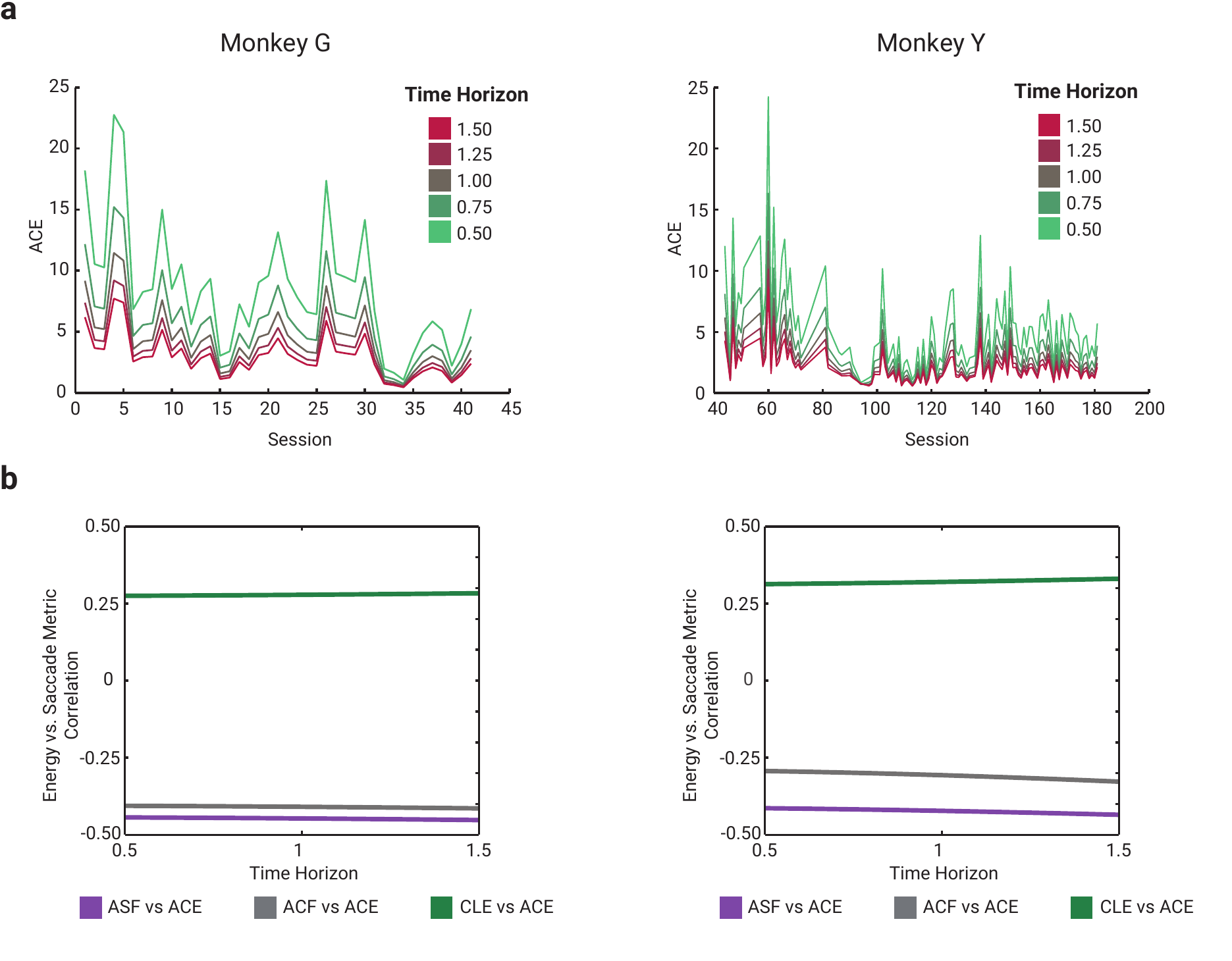}
\captionsetup{labelfont={small, bf}, textfont=small}
\caption{\textbf{Control Energy Dynamics: Time Horizon Parameter Sweep}. \textbf{(a)} Average control energy dynamics across all sessions for Monkey G (left) and Monkey Y (right) using five different values of the time horizon parameter. Smaller time horizon values result in higher magnitude energy values and \emph{vice versa}. The time horizon was set to a value of 1 for all analysis in the main text. \textbf{(b)} All three energy-behavior correlations were re-calculated using the control energy derived from a range of time horizons (0.5 to 1.5 in intervals of 0.1). The change in each energy to behavior (average similarity factor, average complexity factor, and cluster label entropy) correlation is shown as a function of the time horizon (Monkey G - Left; Monkey Y - Right).}
\label{supp_fig6}
\end{figure*}

\newpage
\thispagestyle{fancy}
\begin{figure*}[!t]
\includegraphics[width=\textwidth]{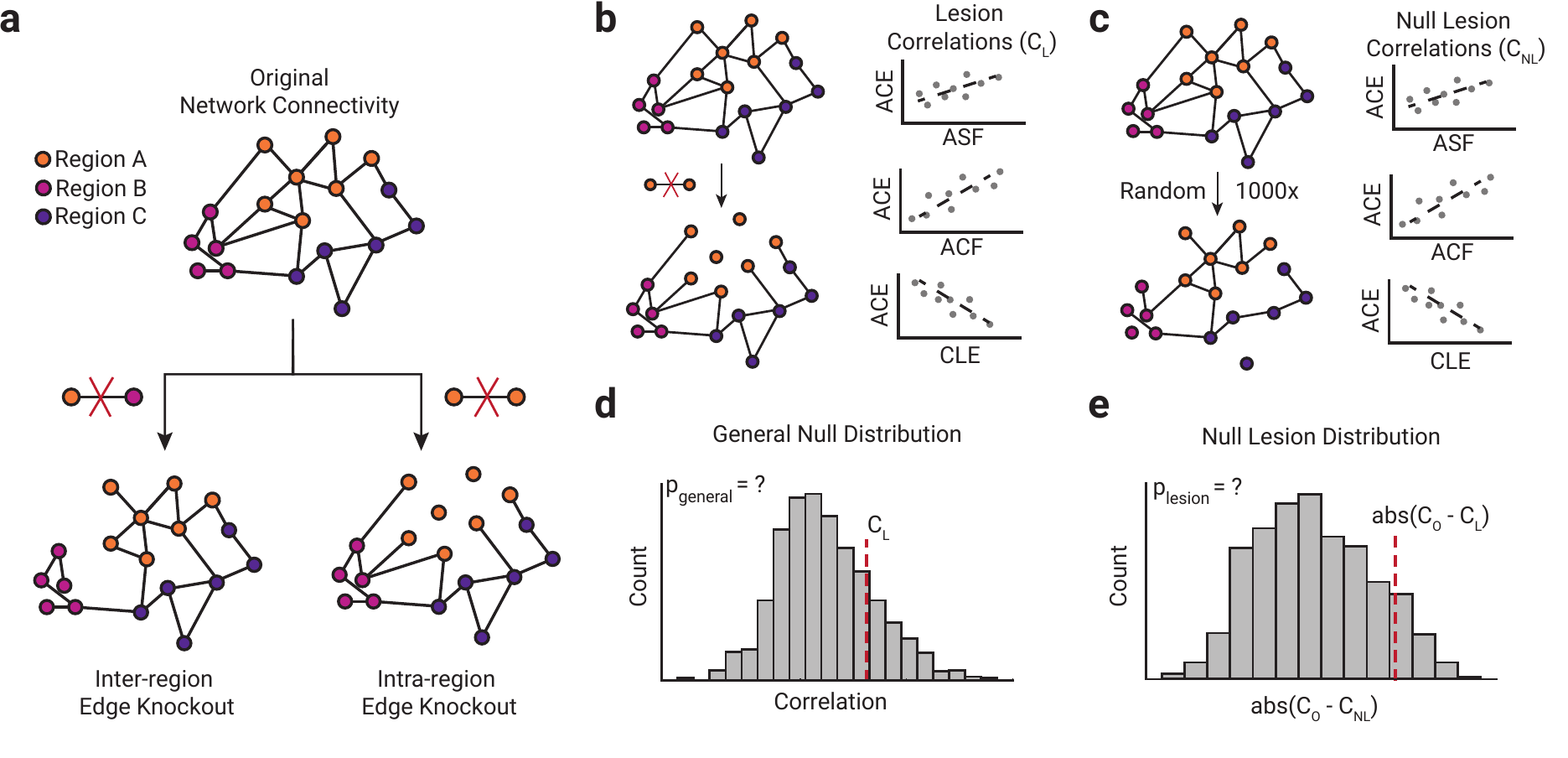}
\captionsetup{labelfont={small, bf}, textfont=small}
\caption{\textbf{Virtual Region Specific Lesion Analysis Guide}. \textbf{(a)} Visual depiction of the lesion analysis workflow. The lesion knockout consists of performing a series of edge-knockouts where (i) all edges between two regions are set to zero in the effective connectivity matrix (inter-region), or where (ii) all edges connecting one region to itself are set to zero (intra-region). \textbf{(b)} For each region specific lesion, the effective connectivity matrix with region specific edges knocked out is used to compute the average control energy and its correlation ($C_{L}$) to the saccade metrics. \textbf{(c)} Random lesions (1000x) were performed to ensure that the lesion-induced disruption of the observed correlations between average control energy and the behavioral metrics was specific to the lesion chosen, and not expected by lesioning the same number of randomly chosen edges. For each random lesion the average control energy and its correlation ($C_{NL}$) to the saccade metrics was computed. \textbf{(d)} The first criterion to test whether the knockout edges were relevant to the observed energy-behavior correlations, was that the region specific lesion resulted in a lesion correlation value, $C_{L}$, that was not significantly different ($p$ > 0.05) from that obtained using the original permutation null model. The obtained $p$-value from such a significance test is referred to as, $p$_value$_{general}$. \textbf{(e)} For the second criterion, a null lesion distribution was created with each value being calculated as the $abs(C_{O} - C_{NL})$, where $C_{O}$ is the observed energy-behavior correlation without any lesions. The significance of the region specific lesion disruption ($abs(C_{O} - C_{L})$) of the observed correlations was determined using a one-tailed test on the null lesion distribution with $\alpha = 0.05$. The obtained $p$-value from such a significance test is referred to as, $p$-value$_{lesion}$.}
\label{supp_fig7}
\end{figure*}

\newpage
\thispagestyle{fancy}
\begin{figure*}[!ht]
\includegraphics[width=\textwidth]{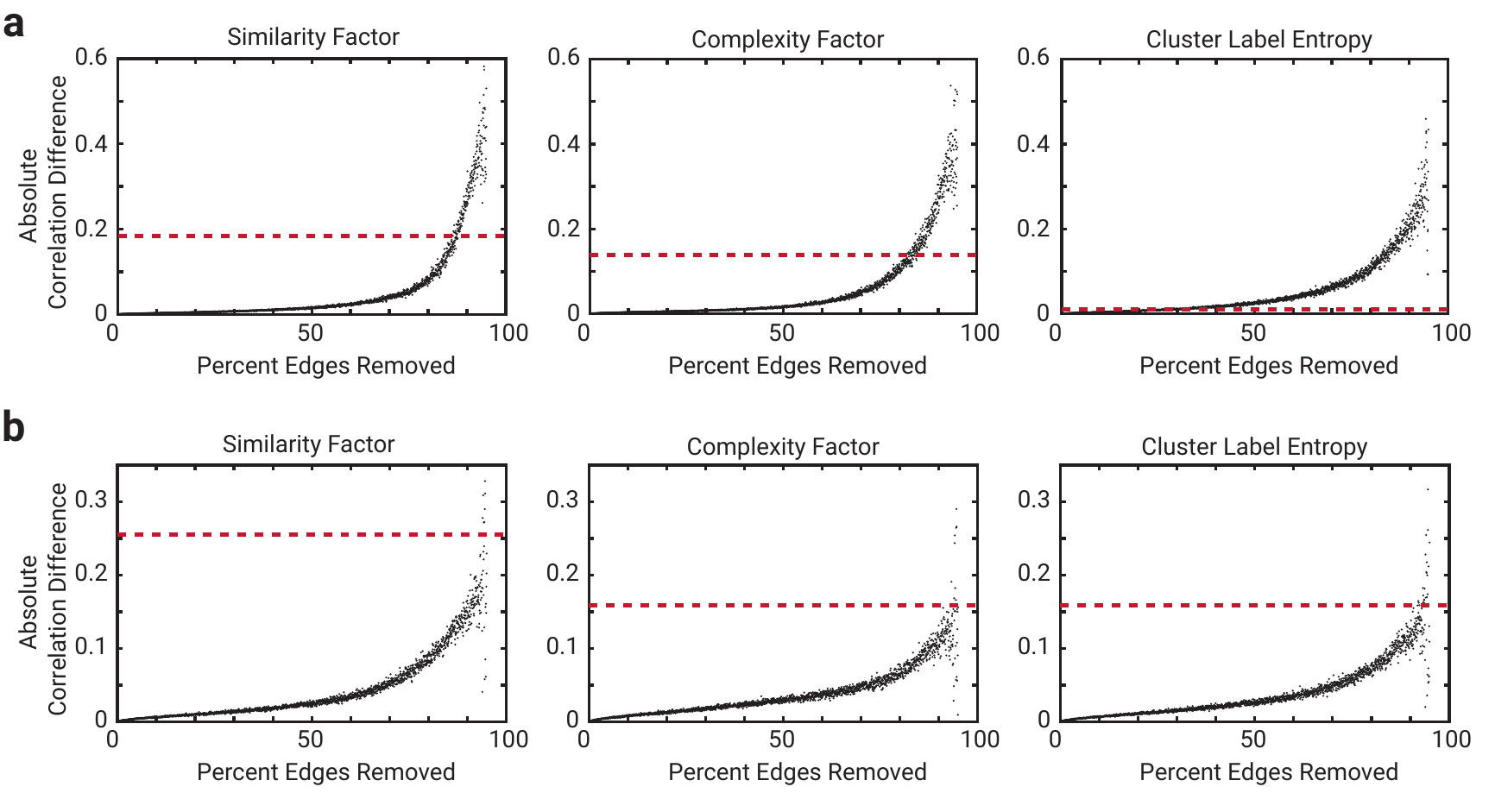}
\captionsetup{labelfont={small, bf}, textfont=small}
\caption{\textbf{Resilience of Energy-Behavior Correlations to Network Disruption}. \textbf{(a-b)} Resilience of energy-behavior correlations as a function of number of edges randomly lesioned for Monkey G \textbf{(a)} and Monkey Y \textbf{(b)}. Every lesion was performed 100 times and each plot value is therefore the average absolute correlation difference. The absolute correlation difference was calculated as the difference between the observed energy-behavior correlation without any lesioning and the correlation after random edges were lesioned from the effective connectivity matrix. Red dashed lines denote the minimum threshold required to significantly disrupt the energy-behavior correlation such that its $p$-value is greater than $\alpha=0.05$.}
\label{supp_fig8}
\end{figure*}